\documentclass[aps,prl,twocolumn,nofootinbib,tightenlines,superscriptaddress,amsmath,amssymb,final,letterpaper]{revtex4-2}

\usepackage[utf8]{inputenc}
\usepackage{calc}
\usepackage{graphicx}
\graphicspath{{../figs/}{figs/}}
\usepackage{amsmath,amssymb,amsthm}
\usepackage{amsfonts}
\usepackage{bbold}
\usepackage{bm}
\usepackage{dsfont}
\usepackage[dvipsnames]{xcolor}
\usepackage{enumitem}
\usepackage{tikz}
\usepackage{float}
\usepackage[percent]{overpic}

\newcommand{\expec}[1]{\langle #1\rangle}

\newcommand{\Var}{\mathrm{Var}}
\newcommand{\Cov}{\mathrm{Cov}}
\newcommand{\CS}{\mathrm{CS}}
\newcommand{\FE}{\mathrm{FE}}

\usepackage[colorlinks=true,linkcolor=blue,urlcolor=blue,citecolor=blue,anchorcolor=blue]{hyperref}

\makeatletter
\g@addto@macro\normalsize{%
  \setlength{\abovedisplayskip}{3pt plus 1pt minus 1pt}%
  \setlength{\belowdisplayskip}{3pt plus 1pt minus 1pt}%
  \setlength{\abovedisplayshortskip}{1pt plus 1pt}%
  \setlength{\belowdisplayshortskip}{1pt plus 1pt}}
\g@addto@macro\small{%
  \setlength{\abovedisplayskip}{3pt plus 1pt minus 1pt}%
  \setlength{\belowdisplayskip}{3pt plus 1pt minus 1pt}%
  \setlength{\abovedisplayshortskip}{1pt plus 1pt}%
  \setlength{\belowdisplayshortskip}{1pt plus 1pt}}
\makeatother

\begin{document}
\title{Detectability limits of scaling laws}

\author{Alec Kirkley}
\email{alec.w.kirkley@gmail.com} 
\affiliation{School of Computing and Data Science, University of Hong Kong, Hong Kong}
\affiliation{Department of Urban Planning and Design, University of Hong Kong, Hong Kong}
\affiliation{Urban Systems Institute, University of Hong Kong, Hong Kong}

\begin{abstract}
Power law scaling relations between size and output are central to quantitative theories of cities, organisms, and other complex systems. Competing theories predict scaling exponents that differ by small fractions, but there is no existing theory for verifying whether a given dataset can even distinguish exponents at the required resolution to address such discrepancies. Here we derive a resolution limit for scaling exponents, giving the smallest exponent difference that any method of analysis can detect. We find that the Hurst exponents governing the evolution of systems' sizes and deviations from the scaling law determine how long a record of growing systems must be before it can separate competing scaling theories. Empirical results suggest that many available data panels are insufficient for reliable scaling model selection. 
\end{abstract}

\maketitle

Across living and social systems, aggregate outputs $Y$ grow as power laws of system size $N$ according to $Y = CN^{\beta}$, where $\beta$ is a scaling exponent and $C$ is a prefactor shifting the amplitude of output. For example, metabolic rate rises as body mass to the power $\beta\approx 3/4$~\cite{Kleiber1932Body,West1997A}, while urban outputs tied to human interaction, such as income and innovation, rise superlinearly with city population ($\beta\approx 1.15$), whereas infrastructure and other quantities that exhibit economies of scale rise sublinearly ($\beta\approx 0.85$)~\cite{Bettencourt2007Growth,Bettencourt2010Deviations,Bettencourt2013The}. The precise value of the scaling exponent is critical, since various proposed mechanisms predict exponents that differ by only $0.05$--$0.1$, including vascular networks~\cite{West1997A}, metabolic partitioning~\cite{Kempes2012Growth}, and metabolic-level boundaries~\cite{Glazier2005Beyond,Glazier2010A} for organisms, and interaction, congestion, and complexity effects for cities~\cite{Ribeiro2017A,Louf2014How,Ribeiro2023Mathematical,GomezLievano2017Explaining}. Claims of universality, meaning a single scaling exponent $\beta$ applies to a whole class of systems~\cite{West2017Scale}, rely on estimation while merging subgroups of systems~\cite{Youn2016Scaling,Shalizi2011Scaling}. Meanwhile, fitted exponents are known to depend on how system boundaries are drawn~\cite{Arcaute2015Constructing,Cottineau2017Diverse}, on the statistical model~\cite{Leitao2016Is,Warton2006Bivariate}, on heavy-tailed sampling artefacts~\cite{GomezLievano2021Artificial}, on the migration of productive people into large cities~\cite{Keuschnigg2019Scaling,Keuschnigg2019Urban,Combes2015Chapter}, and above all on whether the law is fitted across systems at one time or within systems over time~\cite{Bettencourt2020The,Ribeiro2020On,Xu2019Crosssectional,Depersin2018From}. 

These existing scaling debates presuppose that the dataset at hand can distinguish exponents that differ by only small fractions. Information theory can answer such a question independently of the estimate itself, since for a specified stochastic model of the data it gives the smallest parameter difference that any estimator can resolve~\cite{Lehmann1998Theory,Tsybakov2009Introduction}. No such bound has been derived for the stochastic processes~\cite{Bettencourt2020Urban} that govern the deviations of a system from its scaling law, so there is at present no way to determine how many temporal observations are required to reliably resolve competing exponents for a given corpus of systems.

Here we derive a fundamental resolution limit for scaling exponents that depends on the data only through the observed systems' size trajectories and the covariance of their deviations from the scaling law. We find that in panel data with unknown system-specific prefactors, no estimator, biased or unbiased, can distinguish exponents closer than a threshold set by each system's longitudinal size variation weighted by the inverse covariance of its deviations. We then show that the Hurst exponents governing the evolution of system size and deviations from the scaling law determine how quickly one can achieve the resolution limit with respect to the number of observations over time. Empirically, we find that competing scaling laws for income across the corpus of $381$ US metropolitan areas can be distinguished with roughly two decades of panel data by an estimator that attains the resolution limit, whereas a single city would require millennia of observations to answer the same question. We also find notable biases in the cross-sectional exponents for various urban scaling laws, and that mammalian metabolic rate is better summarized with order-specific prefactors than with a universal law.  

\textit{Model.---}We consider panel data for systems $i=1,...,m$ (e.g. organisms, cities) observed at times $t=1,\dots,T$ (e.g. years, decades), with output (e.g. metabolic rate, GDP) $Y_{it}$ and size (e.g. body mass, population) $N_{it}$. Writing $y_{it}=\ln Y_{it}$ and $x_{it}=\ln N_{it}$, a fully general scaling law with heterogeneous prefactors and exponents takes the form
\begin{align}\label{eq:model}
 y_{it} = c_i + \beta_i x_{it} + \xi_{it},
\end{align}
where $c_i=\ln C_i$ and $\beta_i$ are the (log) prefactor and exponent of system $i$, with standard deviations $\sigma_c$ and $\sigma_\beta$ across systems. Subgroups of systems with a common prefactor, such as hot- and cold-blooded organisms~\cite{WEST1999104}, correspond to the distribution of $c$ being a mixture of point masses, whose spread is also accounted for by $\sigma_c$. The deviations $\xi_{it}$ of system $i$ from the scaling law follow a zero-mean stochastic process $f(\bm{\xi})$ whose covariance $\Sigma_{tt'}=\expec{\xi_{it}\xi_{it'}}$ between observation times encodes how long a deviation persists. (Independent observation noise of variance $\sigma_\varepsilon^2$ is the special case $\Sigma_{tt'}=\sigma_\varepsilon^2\delta_{tt'}$.) We take the sizes as strictly exogenous, so $\xi_{it}$ is uncorrelated with $x_{is}$ at all times $s$. This formulation thus excludes feedback from output deviations to growth, such as income-driven migration. One can also include year fixed effects in Eq.~\eqref{eq:model}, which are explored in our downstream empirical analyses.

The questions that a scaling analysis poses, in order of the amount of data they require, are: (1) whether $\beta$ differs from $0$ or from $1$ (whether there is scaling and whether it is nonlinear); (2) whether $\beta$ differs from the value predicted by a competing theory (which is typically $\delta\beta \in [0.05,0.1]$ away); and (3) whether the systems fall into subgroups such that $\sigma_c>0$ or $\sigma_\beta>0$. Treating the prefactors $c_i$ as free is a conservative choice, since (as we show below) no cross-sectional analysis can verify whether $c_i$ covaries with size and the empirical covariance is comparable to the differences between competing theories. The resolution limit we derive can be adapted to address all three questions above.

\textit{The resolution limit.---}Suppose the true scaling exponent is $\beta$ and we ask whether a given panel dataset could rule out a nearby value $\beta+\delta\beta$. Because each system $i$ has its own unknown prefactor $c_i$, the alternative exponent $\beta+\delta\beta$ does not have to explain the data with the same prefactors as the model with exponent $\beta$, and may change each system's prefactor $c_i$ by an amount $a_i\in \mathbb{R}$. For a system $i$ observed at a single size $x_{i0}$, such a shift can offset the change of exponent completely, since $y_{i0}=c_i+\beta x_{i0}+\xi_{i0}=(c_i+a_i)+(\beta+\delta\beta)x_{i0}+\xi_{i0}$ holds for $a_i=-\delta\beta x_{i0}$. Thus, the two models with can produce identical data in the cross-sectional case. However, for a system $i$ observed at several sizes $\{x_{it}\}$, one shift $a_i$ cannot offset the change at all $\{x_{it}\}$, and the mismatch that remains in the models can be detected in the measured panel data. The question of whether $\beta$ can be distinguished from $\beta+\delta\beta$ can therefore be rephrased as how close the model with exponent $\beta+\delta\beta$ comes to the model with exponent $\beta$ when the prefactor shifts $\{a_i\}$ are chosen to align the models as closely as possible. 

Let $\mathcal{P}_0$ and $\mathcal{P}_{\bm{a}}$ represent the model in Eq.~\eqref{eq:model} with parameter configurations $(\{c_i\},\beta)$ and $(\{c_i+a_i\},\beta+\delta\beta)$ respectively. Minimizing the Kullback--Leibler divergence between the two models over the prefactor shifts $\{a_i\}$ gives, to leading order in the separation $\delta\beta$
\begin{align}\label{eq:klmain}
 \min_{\bm{a}}\left\{\mathrm{KL}(\mathcal{P}_{0}\vert\vert \mathcal{P}_{\bm{a}})\right\} = \tfrac{1}{2}\,\mathcal{S}(\bm{x},f)\,\delta\beta^2,
\end{align}
where 
\begin{align}\label{eq:Sdef}
\mathcal{S}(\bm{x},f)=\sum_{i}\tilde{\bm{x}}_i^{\top} \mathcal{I}(f)\, \tilde{\bm{x}}_i     
\end{align}
is the total within-system variation of the system sizes, weighted by the information geometry of the deviations through the Fisher information matrix $\mathcal{I}(f)$ of the joint deviation density $f(\bm{\xi})$, and $\tilde x_{it}=x_{it}-\bm{1}^{\top}\mathcal{I}(f)\bm{x}_i/\bm{1}^{\top}\mathcal{I}(f)\bm{1}$. We will call $\mathcal{S}(\bm{x},f)$ the covariance-adjusted size fluctuations. Intuitively, a prefactor shift $a_i$ absorbs an exponent change wherever the size $x_{it}$ of a system does not change, so only changes of size are informative, and a change of size carries less information when the deviations are more strongly correlated over the same time interval. For Gaussian deviations of covariance $\bm{\Sigma}$ the Fisher information matrix is $\mathcal{I}(f)=\bm{\Sigma}^{-1}$, and $\tilde{\bm{x}}_i$ is demeaned in the metric of $\bm{\Sigma}^{-1}$ (the ordinary mean when the deviations are independent). Among all deviation processes of a given covariance $\bm{\Sigma}$, the Gaussian carries the least information ($\mathcal{I}\succeq\bm{\Sigma}^{-1}$ with equality only for the Gaussian), so evaluating $\mathcal{S}(\bm{x},f)$ with $\mathcal{I}=\bm{\Sigma}^{-1}$ gives the smallest information and hence the largest, most conservative resolution limit for any estimator. We will thus adopt this evaluation, $\mathcal{S}(\bm{x},\bm{\Sigma})=\sum_i\tilde{\bm{x}}_i^{\top}\bm{\Sigma}^{-1}\tilde{\bm{x}}_i$, throughout for downstream results.

To resolve two exponents $\beta$ and $\beta+\delta\beta$ at a confidence of $z$ standard errors, the two exponents must differ by at least $z$ times the smallest standard error attainable, given by the Cram\'er--Rao bound. This variance bound is attained by the generalized least-squares fit with the prefactors eliminated. Meanwhile, by Eq.~\eqref{eq:klmain}, this separation is where the minimized KL divergence reaches $z^2/2$, which is a property of the two data distributions rather than any single estimator, implying that no procedure, biased or unbiased, can resolve a smaller separation~\cite{Tsybakov2009Introduction}. Putting this all together we have an estimator-independent \emph{resolution limit}
\begin{align}\label{eq:resolution}
 \delta\beta_{\min} = z/\sqrt{\mathcal{S}(\bm{x},f)},
\end{align}
with $z$ the desired confidence in standard errors. Here we adopt the conventional $z=2$, so that $\delta\beta_{\min}=2/\sqrt{\mathcal{S}(\bm{x},\bm{\Sigma})}$. For independent noise of density $f$, $\mathcal{S}(\bm{x},f)=J_fV(\bm{x})$ with $V(\bm{x})=\sum_{it}(x_{it}-\bar x_i)^2$ the total within-system variation of log size and $J_f=\int f'(\xi)^2/f(\xi)\,d\xi$ the (scalar) Fisher information of the noise density, which the Gaussian minimizes at $\sigma_\varepsilon^{-2}$ for independent observation noise $\Sigma_{tt'}=\sigma_\varepsilon^2\delta_{tt'}$. Formal derivations are given in SM Sec.~\ref{app:KL}.

In a cross-sectional snapshot, every system $i$ is observed at a single size $x_{i0}$, so $\mathcal{S}=0$ for every $\bm{\Sigma}$ and, with free prefactors, no analysis of a cross section can distinguish any two exponents reliably. In particular, a snapshot cannot establish that a relation is nonlinear ($\beta\neq 1$), which is the question at the center of the urban scaling debate~\cite{Leitao2016Is,Bettencourt2007Growth}. The two estimators commonly used by scaling studies illustrate the issue. The cross-sectional estimator $\hat\beta_{\CS}$ is the least-squares slope across systems at one time, and the within (fixed-effects) estimator $\hat\beta_{\FE}$ is the least-squares slope after each system's record is centered on its own mean, which removes the prefactors. Formally,
\begin{align}
 \hat\beta_{\CS}&=\frac{\sum_i(x_{i0}-\bar x_0)(y_{i0}-\bar y_0)}{\sum_i(x_{i0}-\bar x_0)^2},\label{eq:CS}\\
 \hat\beta_{\FE}&=\frac{\sum_{it}(x_{it}-\bar x_i)(y_{it}-\bar y_i)}{V(\bm{x})},\label{eq:FE}
\end{align}
where $\bar x_0$ and $\bar y_0$ are means across systems and $\bar x_i$ and $\bar y_i$ are means over time for system $i$. The within estimator converges to the within-system response $\beta$ and attains the limit of Eq.~\eqref{eq:resolution} for independent deviations (SM Sec.~\ref{app:KL}), while a single snapshot only estimates the sum~\cite{Mundlak1978On,Chamberlain1982Multivariate}
\begin{align}\label{eq:csbias}
 \hat\beta_{\CS} \;\longrightarrow\; \beta + b,
 ~~~
 b=\frac{\Cov(c,x)}{\Var(x)}.
\end{align}
The assumption $b=0$ that underlies every cross-sectional exponent cannot be verified from the snapshot itself, and the urban panel analyses below estimate $b$ to be similar to the differences between competing theories, so this assumption does not always hold up in practice~\cite{Bettencourt2020The,Ribeiro2020On,Xu2019Crosssectional,Depersin2018From}.

\textit{Two dynamical regimes.---}Once systems are observed over time, $\mathcal{S}$ becomes non-zero and grows with the observation window at a rate set by the Hurst exponents of the sizes and deviations, defined by the mean-square displacements $\langle(x_{i,t+\ell}-x_{it})^2\rangle\propto\ell^{2H_x}$ and $\langle(\xi_{i,t+\ell}-\xi_{it})^2\rangle\propto\ell^{2H_\xi}$, where $H=1/2$ is ordinary diffusion~\cite{Metzler2000The}. Two distinct dynamical regimes for detectability in longitudinal panels are determined by whether the deviations are stationary or have stationary increments (a random walk, $H_\xi=1/2$, being the simplest case).

\begin{figure}[t]
    \centering
    \includegraphics[width=\columnwidth]{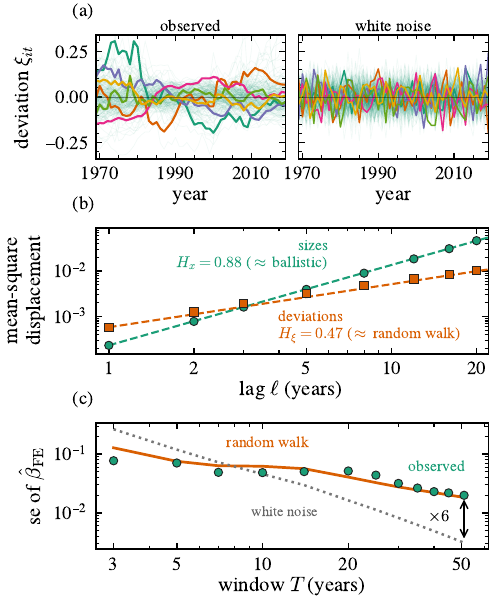}
    \caption{
    \textbf{City sizes, scaling deviations, and exponent precision for US metropolitan areas.}
    (a)~Deviations $\xi_{it}$ of log income from the scaling law for $381$ US metropolitan areas over the period 1969--2019 (six cities highlighted), compared with white noise of the same variance.
    (b)~Mean-square displacement versus lag for log population (circles) and the scaling deviations of panel~(a) (squares).
    (c)~Standard error of the within-city income exponent $\hat\beta_{\FE}$ versus window length (windows ending in 2019), against its prediction for random-walk deviations [SM Eq.~\eqref{eq:varFE}] (solid) and for white noise [$\sigma_\varepsilon/\sqrt{V}$, Eq.~\eqref{eq:Sstat} with $\tau=0$] (dotted). Both curves use the actual size trajectories with no parameter tuning.
    }
    \label{fig:dynamics}
\end{figure}

If the deviations are stationary, the mean-square displacement saturates such that $H_\xi\to0$ at large lags. In this case, with marginal variance $\sigma_\xi^2$ and correlation time $\tau$, we have to leading order in large $T$
\begin{align}\label{eq:Sstat}
 \mathcal{S}\simeq\frac{m\,c_G\,T^{1+2H_x}}{\sigma_\xi^2(1+2\tau)},
\end{align}
where $c_G$ is a constant that depends on the size dynamics (SM Sec.~\ref{app:Hurst}). Steady growth at rate $g$ is ballistic ($H_x=1$) and gives the $T^{3}$ law of longitudinal design~\cite{Willett1989Some,Raudenbush2001Effects}, Gibrat-type multiplicative growth~\cite{Sutton1997Gibrats,West1988Asymptotic} is diffusive ($H_x=1/2$) and gives a $T^{2}$ law, and sizes that fluctuate about a steady state, as for adult organisms or saturated cities, are confined ($H_x=0$) and scale with~$T$.

If the deviations have stationary increments (nonstationary, $0<H_\xi<1$), we have instead
\begin{align}\label{eq:Snonstat}
 \mathcal{S}\simeq\frac{m}{\sigma_\eta^2}\left[g^2\,T^{2(1-H_\xi)}+\sigma_g^2\,T\right],
\end{align}
where $\sigma_\eta^2$ is the marginal variance of the increments $\Delta\xi_{it}$, and $g$ and $\sigma_g$ are the mean and standard deviation of the size increments $\Delta x_{it}$. (This expression is asymptotically exact for $H_\xi=1/2$, and $O(1)$ prefactors that arise for $H_\xi\neq1/2$ are given in SM Sec.~\ref{app:Hurst}.) The result does not depend on $H_x$, since differencing reduces each time step to a single increment regardless of the size dynamics.

Inverting Eq.~\eqref{eq:resolution} in each regime gives the record length $T^{*}$ needed to resolve a separation $\delta\beta$, such that
\begin{align}\label{eq:tstar}
 \text{stationary}&:~~T^{*}=\left(\frac{4\sigma_\xi^2(1+2\tau)}{c_G\, m\,\delta\beta^2}\right)^{\!\frac{1}{1+2H_x}},\\
 \text{nonstationary}&: ~~g^2\,T^{*\,2(1-H_\xi)}+\sigma_g^2\,T^{*}=\frac{4\sigma_\eta^2}{m\,\delta\beta^2}.\nonumber
\end{align}
The nonstationary case is linear at $H_\xi=1/2$, with $T^{*}=4\sigma_\eta^2/[(g^2+\sigma_g^2)m\,\delta\beta^2]$. For general $H_\xi$, it reduces to $T^{*}=[4\sigma_\eta^2/(g^2m\,\delta\beta^2)]^{1/[2(1-H_\xi)]}$ when drift dominates the size increments and to $T^{*}=4\sigma_\eta^2/(\sigma_g^2m\,\delta\beta^2)$ when the Gibrat steps dominate. 

\begin{figure}
    \centering
    \includegraphics[width=\columnwidth]{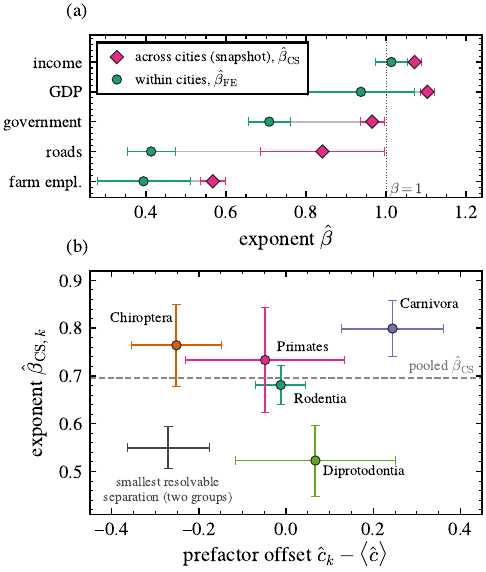}
    \caption{
    \textbf{Urban scaling exponents and mammalian metabolic subgroups.}
    (a)~Cross-sectional ($\hat\beta_{\CS}$) and within-city ($\hat\beta_{\FE}$) exponents for five urban outputs. Snapshot error bars are $\pm 2$ standard deviations across the per-timestep cross-sectional fits, while within-city error bars are $2\sigma$ intervals clustered by city.
    (b)~Prefactor offset and exponent for basal metabolic rate versus body mass~\cite{Jones2009PanTHERIA} in five mammalian orders ($2\sigma$ crosshairs), along with the pooled exponent $0.697\pm 0.014$ (dashed). Using the subgroup limits of Eq.~\eqref{eq:sigbetamin}, the $2\sigma_{c,\min}$ threshold and $2\sigma_{\beta,\min}$ threshold are indicated horizontally and vertically respectively by the grey cross. Intervals and thresholds are unclustered (SM Sec.~\ref{app:data} shows the result when clustering residuals by family).
    }
    \label{fig:empirical}
\end{figure}

\textit{Cities.---}We build balanced annual panels of $m=381$ US metropolitan areas over $T=51$ years (1969--2019, with fixed city boundaries) for personal income (superlinear) and farm employment (sublinear), with national trends removed as year effects (SM Sec.~\ref{app:data}). Figure~\ref{fig:dynamics}(a) shows that each city's deviation from the scaling law is correlated over decades, unlike white noise of the same variance. Figure~\ref{fig:dynamics}(b) shows the two Hurst exponents of Eqs.~\eqref{eq:Sstat}--\eqref{eq:Snonstat}, with sizes being nearly ballistic ($H_x=0.88$)---possibly reflecting migration-driven drift~\cite{Verbavatz2020The}---and deviations diffusing ($H_\xi=0.47$, slightly below $1/2$ because a small stationary component adds a constant to the mean-square displacement) with nearly uncorrelated yearly steps (SM Sec.~\ref{app:deviations}). Cities therefore fall in the nonstationary regime for this corpus. In Fig.~\ref{fig:dynamics}(c) the standard error of $\hat\beta_{\FE}$ predicted for random-walk deviations ($H_\xi=1/2$), evaluated on the actual size trajectories with the measured step variances and no free parameter (SM Sec.~\ref{app:deviations}), follows the observed uncertainty at every window from $5$ to $51$ years (SM Table~\ref{tab:tracking}), whereas the white noise law [Eq.~\eqref{eq:Sstat} with $\tau=0$] overstates the precision by a factor of $6$ at $T=51$.

\begin{figure}[b]
    \centering
    \includegraphics[width=\columnwidth]{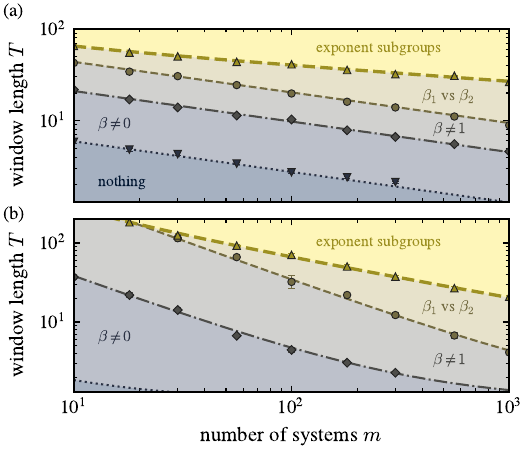}
    \caption{
    \textbf{Resolution limit boundaries in the $(m,T)$ plane.}
    (a)~Resolution limits for ballistic ($H_x=1$) growth and white noise deviations (stationary, $\tau=0$). The parameters $\{\tilde g,\sigma_\xi\}=\{0.010,0.066\}$ are measured from the US metropolitan income panel (SM Sec.~\ref{app:deviations}) and we set $\delta\beta\in\{1,0.15,0.05\}$ for the three exponent location questions $\{\beta\neq 0,\beta\neq 1,\beta_1\neq \beta_2\}$ respectively, while letting $\sigma_\beta=0.05$ be the scale that separates distinct exponent subgroups. Theoretical boundaries from Eqs.~\eqref{eq:tstar}--\eqref{eq:sigbetamin2} are shown as dashed and dotted curves, and the windows at which each question is resolved in half of $200$ simulated panels per point (SM Sec.~\ref{app:synthetic}) as markers.
    (b)~Same boundaries for random-walk deviations (nonstationary, $H_\xi=1/2$) at the calibration $\{\tilde g,\sigma_g,\sigma_\eta^2\}=\{0.010,0.012,5.1\times10^{-4}\}$ measured from the same panel (SM Sec.~\ref{app:deviations}), with the same $\delta\beta$ values and $\sigma_\beta=0.1$. 
    }
    \label{fig:phase}
\end{figure}

Under the random-walk law for the deviations, the required record lengths for distinguishing common scaling theories can be computed for this corpus (SM Table~\ref{tab:ladder}). For this set of cities, only a few years of data are sufficient to establish that income scales with population and resolve whether the scaling is nonlinear ($\beta\neq 0,1$), whereas competing superlinear theories falling $\delta\beta\sim 0.05$ apart~\cite{Ribeiro2017A,Louf2014How,GomezLievano2017Explaining} would require about two decades of measurements for an estimator achieving the resolution limit and nearly four decades for the standard within-city (fixed effects) estimator, which does not attain the limit in this nonstationary regime. (The analytical boundary of Fig.~\ref{fig:phase}(b) gives a shorter window, see SM Sec.~\ref{app:deviations}.) Few existing panels reach this length, and smaller corpora need even longer. Meanwhile, a single city's exponent cannot be reliably measured on any historical record, since the random-walk law extrapolated to $m=1$ gives windows of centuries to millennia for this dataset (SM Sec.~\ref{app:deviations}).

Across the US cities in the corpus, income scales superlinearly with population, $\hat\beta_{\CS}=1.071\pm 0.018$ ($2\sigma$ intervals throughout). Within cities, the same panel gives $\hat\beta_{\FE}=1.013\pm 0.040$, which is below the cross-sectional value and consistent with simple proportionality [Fig.~\ref{fig:empirical}(a)]. The gap between the two estimates the bias term of Eq.~\eqref{eq:csbias} to be $b\approx 0.06$, consistent with the positive correlation we would expect for productive workers and activities sorting into large cities~\cite{Keuschnigg2019Urban,Combes2015Chapter}. A positive bias $b$ appears for every indicator we examine, as the snapshot exponent exceeds the within-city exponent for income, GDP, government employment, farm employment, and road length, which cover both sub- and superlinear scaling regimes [Fig.~\ref{fig:empirical}(a), SM Sec.~\ref{app:data}, Fig.~\ref{fig:other}]~\cite{Burghardt2022Road,Burghardt2024Analyzing}. 

\textit{Universality and subgroups.---}We can use the same KL divergence criterion to resolve subgroup structure, indicated by $\sigma_c>0$ or $\sigma_\beta>0$. For stationary deviations, we have
\begin{align}\label{eq:sigbetamin}
\sigma_{\beta,\min}\sim T^{-1/2-H_x}m^{-1/4},~~
\sigma_{c,\min}\sim T^{-1/2}m^{-1/4},
\end{align}
while for nonstationary deviations we have
\begin{align}\label{eq:sigbetamin2}
\sigma_{\beta,\min}\sim T^{-(1-H_\xi)}m^{-1/4}
\end{align}
for drift-dominated size increments. (Gibrat-dominated increments give $T^{-1/2}m^{-1/4}$ for every $H_\xi$.) Meanwhile, prefactor structure for nonstationary deviations does not involve the whole time window $T$. Both limits are attained by an excess scatter test (SM Sec. E). Figure~\ref{fig:empirical}(b) shows the stationary case for mammalian metabolism~\cite{Jones2009PanTHERIA}, in which the estimated prefactor spread of five orders exceeds $\sigma_{c,\min}$ by a factor of roughly two, while their exponents show no excess scatter beyond $\sigma_{\beta,\min}$ once residuals are clustered by family [Fig.~\ref{fig:empirical}(b), SM Sec.~\ref{app:data}]. 

Figure~\ref{fig:phase} draws the boundaries of Eqs.~\eqref{eq:tstar}--\eqref{eq:sigbetamin2} in the $(m,T)$ plane for both the stationary (top) and nonstationary (bottom) regimes, each giving the boundary at which a two standard error test resolves the question in half of the simulations. In the nonstationary regime [Fig.~\ref{fig:phase}(b)] the location boundaries steepen from $m^{-1/3}$ to $m^{-1}$, so corpus size is far more important than in the stationary regime.

\textit{Conclusion.---}Here we have derived a set of information theoretic resolution limits that determine when a dataset can support various scaling claims, along with the mechanisms by which the dynamics of systems' sizes and scaling deviations determine these limits. Natural extensions include resolution limits for mixtures of mechanisms, deviation processes with feedback mechanisms for system growth, and clustering methods that automatically extract a single law whenever the data lie below the subgroup limit. 

\textit{Data and code availability.---}All code and data to reproduce the results are available at \url{https://github.com/aleckirkley/scaling-detectability}.


\clearpage
\onecolumngrid
\raggedbottom
\appendix

\setcounter{figure}{0}
\setcounter{table}{0}
\renewcommand{\thefigure}{A\arabic{figure}}
\renewcommand{\thetable}{A\arabic{table}}
\setcounter{secnumdepth}{3} 

\begin{center}
  \textbf{\large Supplemental Material for: \\ \vspace{0.25cm}
Detectability limits of scaling laws\vspace{0.25cm}} \\[.2cm]
  Alec Kirkley$^{1,2,3}$ \\ [.1cm]
  {\itshape ${}^1$School of Computing and Data Science, University of Hong Kong, Hong Kong \\
  ${}^2$Department of Urban Planning and Design, University of Hong Kong, Hong Kong \\
  ${}^3$Urban Systems Institute, University of Hong Kong, Hong Kong
   }
\end{center}

\section{Derivation of the resolution limit and related quantities}\label{app:KL}

\subsection{Profiled KL divergence}

Let the deviations of system $i$ have joint density $f(\bm{\xi}_i)$ with covariance $\bm{\Sigma}$, where $f$ does not depend on $(\{c_i\},\beta)$ (the standard assumption of Eq.~\eqref{eq:model}). Deviations $\bm{\xi}$ whose distribution changes with the model parameters would carry additional information about $\beta$, so would naturally lower the associated resolution limit. Under $\mathcal{P}_0$ the mean of $y_{it}$ is $\mu_{it}=c_i+\beta x_{it}$, while under $\mathcal{P}_{\bm{a}}$ it is $(c_i+a_i)+(\beta+\delta\beta)\,x_{it}$, so the mean shifts by $\delta\beta\,x_{it}+a_i$. Because the parameters enter Eq.~\eqref{eq:model} only through the mean, the density of $\bm{y}_i$ under any parameter configuration is the fixed deviation density $f$ shifted to that mean, $f(\bm{y}_i-\bm{\mu}_i)$, and a change of parameters shifts only the mean vector. 

We will let $\bm{\delta}_i=\bm{\mu}'_i-\bm{\mu}_i=\delta\beta\,\bm{x}_i+a_i\bm{1}$ for the mean shift of system $i$ between $\mathcal{P}_0$ and $\mathcal{P}_{\bm{a}}$. Since systems are independent, the divergence is a sum over systems, and for one system
\begin{align}
 \mathrm{KL}_i(\bm{\delta}_i)
 =\int f(\bm{\xi})\,\ln\frac{f(\bm{\xi})}{f(\bm{\xi}-\bm{\delta}_i)}\,d\bm{\xi}.
\end{align}
Expanding $\ln f(\bm{\xi}-\bm{\delta}_i)$ to second order in $\bm{\delta}_i$,
\begin{align}
 \ln f(\bm{\xi}-\bm{\delta}_i)
 =\ln f(\bm{\xi})-\bm{\delta}_i^\top\nabla\ln f(\bm{\xi})
 +\tfrac{1}{2}\bm{\delta}_i^\top\nabla\nabla^\top\!\ln f(\bm{\xi})\,\bm{\delta}_i+O(|\bm{\delta}_i|^3),
\end{align}
and taking the expectation under $f$, the first-order term vanishes because $\int f\,\nabla\ln f\,d\bm{\xi}=\int\nabla f\,d\bm{\xi}=\nabla\int f\,d\bm{\xi}=0$. Cancelling the leading order terms $\int f \ln f d\bm{\xi}$, the second-order term gives a system-level KL divergence of
\begin{align}
 \mathrm{KL}_i(\bm{\delta}_i)
 =-\tfrac{1}{2}\bm{\delta}_i^\top\Big\langle\nabla\nabla^\top\!\ln f\Big\rangle_f\bm{\delta}_i+O(|\bm{\delta}_i|^3)
 =\tfrac{1}{2}\bm{\delta}_i^\top\mathcal{I}(f)\,\bm{\delta}_i+O(|\bm{\delta}_i|^3),
\end{align}
where
\begin{align}
\mathcal{I}(f)=\langle\nabla\ln f\,\nabla^\top\!\ln f\rangle_f=-\langle\nabla\nabla^\top\!\ln f\rangle_f    
\end{align}
is the $T\times T$ Fisher information matrix of the deviation process. The two forms of $\mathcal{I}(f)$ agree because $\partial_s\partial_t\ln f=\partial_s\partial_t f/f-\partial_s\ln f\,\partial_t\ln f$, and $\int\partial_s\partial_t f\,d\bm{\xi}=\partial_s\partial_t\int f\,d\bm{\xi}=0$ for a density that decays at infinity, so $-\langle\partial_s\partial_t\ln f\rangle_f=\langle\partial_s\ln f\,\partial_t\ln f\rangle_f$. Summing over systems, we have
\begin{align}
 \mathrm{KL}(\mathcal{P}_0\,\|\,\mathcal{P}_{\bm{a}})
 =\frac{1}{2}\sum_i\bm{\delta}_i^\top\mathcal{I}(f)\,\bm{\delta}_i+O(|\bm{\delta}|^3),
 \qquad
 \bm{\delta}_i=\delta\beta\,\bm{x}_i+a_i\bm{1}.
\end{align}
The expansion is exact at all orders for Gaussian $f$, for which $\ln f$ is quadratic and $\mathcal{I}(f)=\bm{\Sigma}^{-1}$. For any other $f$ the neglected terms are $O(\delta\beta^3)$ and do not affect the resolution limit, which is defined by the quadratic coefficient. Setting $\partial_{a_i}\mathrm{KL}=0$ gives
\begin{align}
 a_i^*=-\delta\beta\,\frac{\bm{1}^\top\mathcal{I}(f)\,\bm{x}_i}{\bm{1}^\top\mathcal{I}(f)\,\bm{1}}.
\end{align}
Defining $\tilde{\bm{x}}_i=\bm{x}_i-\frac{\bm{1}^\top\mathcal{I}(f)\,\bm{x}_i}{\bm{1}^\top\mathcal{I}(f)\,\bm{1}}\,\bm{1}$, so that $\bm{1}^\top\mathcal{I}(f)\,\tilde{\bm{x}}_i=0$, the profiled KL divergence is
\begin{align}\label{eq:klgeneral}
 \min_{\bm{a}}\mathrm{KL}(\mathcal{P}_0\,\|\,\mathcal{P}_{\bm{a}})
 = \frac{\delta\beta^2}{2}\sum_i\tilde{\bm{x}}_i^\top\mathcal{I}(f)\,\tilde{\bm{x}}_i
 = \frac{\delta\beta^2}{2}\,\mathcal{S}(\bm{x},f),
\end{align}
which is Eq.~\eqref{eq:klmain}. For Gaussian deviations $\mathcal{I}(f)=\bm{\Sigma}^{-1}$ and the expansion is exact at all orders, so
\begin{align}\label{eq:klcorr}
 \min_{\bm{a}}\mathrm{KL}(\mathcal{P}_0\,\|\,\mathcal{P}_{\bm{a}})
 = \frac{\delta\beta^2}{2}\sum_i\tilde{\bm{x}}_i^\top\bm{\Sigma}^{-1}\tilde{\bm{x}}_i
 = \frac{\delta\beta^2}{2}\,\mathcal{S}(\bm{x},\bm{\Sigma}),
\end{align}
with $\tilde{\bm{x}}_i=\bm{x}_i-\frac{\bm{1}^\top\bm{\Sigma}^{-1}\bm{x}_i}{\bm{1}^\top\bm{\Sigma}^{-1}\bm{1}}\,\bm{1}$. Independent noise ($\bm{\Sigma}=\sigma_\varepsilon^2\bm{I}$) further reduces $\tilde{\bm{x}}_i$ to the ordinary demeaned vector $\bm{x}_i-\bar x_i\bm{1}$ and gives $\mathcal{S}=V(\bm{x})/\sigma_\varepsilon^2$.

The Gaussian is the worst case for resolving the exponents, since among all densities with covariance $\bm{\Sigma}$, the Gaussian minimizes the Fisher information, $\mathcal{I}(f)\succeq\bm{\Sigma}^{-1}$, with equality only for the Gaussian~\cite{Cover2006Elements}. The Gaussian evaluation $\mathcal{S}(\bm{x},\bm{\Sigma})$ is therefore a lower bound on $\mathcal{S}(\bm{x},f)$ for any $f$ with covariance $\bm{\Sigma}$, so we can adopt it throughout the paper as the conservative evaluation.

\subsection{Attainment of the variance bound by the Generalized Least-Squares (GLS) estimator}

The model for system~$i$ is $\bm{y}_i=c_i\bm{1}+\beta\bm{x}_i+\bm{\xi}_i$ with $\mathrm{Cov}(\bm{\xi}_i)=\bm{\Sigma}$, and the generalized least-squares (GLS) estimator minimizes the weighted residual
\begin{align}
 L(\beta,c_1,\dots,c_m)=\sum_i\left(\bm{y}_i-c_i\bm{1}-\beta\bm{x}_i\right)^\top\bm{\Sigma}^{-1}\left(\bm{y}_i-c_i\bm{1}-\beta\bm{x}_i\right)
\end{align}
over $\beta$ and the $m$ prefactors. At fixed $\beta$, setting $\partial L/\partial c_i=0$ gives
\begin{align}
 \hat c_i(\beta)=\frac{\bm{1}^\top\bm{\Sigma}^{-1}(\bm{y}_i-\beta\bm{x}_i)}{\bm{1}^\top\bm{\Sigma}^{-1}\bm{1}},
\end{align}
and substituting back, we have $\bm{y}_i-\hat c_i(\beta)\bm{1}-\beta\bm{x}_i=\tilde{\bm{y}}_i-\beta\tilde{\bm{x}}_i$, where $\tilde{\bm{u}}=\bm{u}-\bm{1}\,(\bm{1}^\top\bm{\Sigma}^{-1}\bm{u})/(\bm{1}^\top\bm{\Sigma}^{-1}\bm{1})$ removes the weighted mean of any vector $\bm{u}$, so that $\bm{1}^\top\bm{\Sigma}^{-1}\tilde{\bm{u}}=0$, and $\tilde{\bm{x}}_i$ is the vector of Eq.~\eqref{eq:Sdef}. The profiled objective is therefore
\begin{align}
 L_{\mathrm{p}}(\beta)=\sum_i\left(\tilde{\bm{y}}_i-\beta\tilde{\bm{x}}_i\right)^\top\bm{\Sigma}^{-1}\left(\tilde{\bm{y}}_i-\beta\tilde{\bm{x}}_i\right),
\end{align}
which is a quadratic in $\beta$ whose minimizer is
\begin{align}
 \hat\beta_{\mathrm{GLS}}
 =\frac{\sum_i\tilde{\bm{x}}_i^\top\bm{\Sigma}^{-1}\tilde{\bm{y}}_i}{\sum_i\tilde{\bm{x}}_i^\top\bm{\Sigma}^{-1}\tilde{\bm{x}}_i}
 =\frac{\sum_i\tilde{\bm{x}}_i^\top\bm{\Sigma}^{-1}\bm{y}_i}{\mathcal{S}(\bm{x},\bm{\Sigma})}.
\end{align}
Inserting the model for $\bm{y}_i$ and using the same orthogonality to drop the $c_i\bm{1}$ terms and to write $\tilde{\bm{x}}_i^\top\bm{\Sigma}^{-1}\bm{x}_i=\tilde{\bm{x}}_i^\top\bm{\Sigma}^{-1}\tilde{\bm{x}}_i$, we have
\begin{align}
 \hat\beta_{\mathrm{GLS}}=\beta+\frac{\sum_i\tilde{\bm{x}}_i^\top\bm{\Sigma}^{-1}\bm{\xi}_i}{\mathcal{S}(\bm{x},\bm{\Sigma})},
\end{align}
which is unbiased conditional on the sizes, and linear in the deviations. Its variance follows from $\mathrm{Cov}(\bm{\xi}_i)=\bm{\Sigma}$ and the independence of systems, thus
\begin{align}
 \mathrm{Var}(\hat\beta_{\mathrm{GLS}})
 =\frac{\sum_i\tilde{\bm{x}}_i^\top\bm{\Sigma}^{-1}\,\bm{\Sigma}\,\bm{\Sigma}^{-1}\tilde{\bm{x}}_i}{\mathcal{S}(\bm{x},\bm{\Sigma})^2}
 =\frac{\sum_i\tilde{\bm{x}}_i^\top\bm{\Sigma}^{-1}\tilde{\bm{x}}_i}{\mathcal{S}(\bm{x},\bm{\Sigma})^2}
 =\frac{1}{\mathcal{S}(\bm{x},\bm{\Sigma})},
\end{align}
which equals the Cram\'er--Rao bound $1/\mathcal{S}$ of Eq.~\eqref{eq:resolution}. Only the covariance of the deviations was needed here, rather than their full distribution, so the GLS estimator attains the precision $1/\mathcal{S}$ for any noise of covariance $\bm{\Sigma}$. Together with the impossibility result of Eq.~\eqref{eq:resolution}, this makes $\delta\beta_{\min}$ an exact resolution limit for the data, in that no procedure can resolve a smaller exponent separation and the standard GLS procedure attains the limit. For independent noise, $\bm{\Sigma}=\sigma_\varepsilon^2\bm{I}$, the weighted mean is the ordinary mean, $\hat\beta_{\mathrm{GLS}}=\hat\beta_{\FE}$ is the within estimator of Eq.~\eqref{eq:FE}, and $\mathrm{Var}(\hat\beta_{\FE})=\sigma_\varepsilon^2/V(\bm{x})$.

\subsection{Connection between KL threshold and confidence level}

Setting $\delta\beta=z/\sqrt{\mathcal{S}}$ in Eq.~\eqref{eq:klmain} gives $\min_{\bm{a}}\mathrm{KL}=z^2/2$, so the resolution limit at $z$ standard errors is the separation at which the divergence between the two closest possible data distributions reaches $z^2/2$. This is a statement about the distributions rather than any estimator, since for arbitrary distributions the summed type-I and type-II error probabilities of any test are at least $\tfrac{1}{2}e^{-\mathrm{KL}}$~\cite{Tsybakov2009Introduction}. We adopt $\mathrm{KL}=z^2/2$ as the convention defining a separation of $z$ standard errors, and the bound $\tfrac{1}{2}e^{-\mathrm{KL}}$ guarantees that below it no test, biased or unbiased, distinguishes $\mathcal{P}_0$ from $\mathcal{P}_{\bm{a}}$ with small error.

\subsection{$\mathcal{S}=J_fV$ for independent noise}

When the deviations $\bm{\xi}$ are independent with common density $f$ (not necessarily Gaussian), the Fisher information matrix is diagonal, $\mathcal{I}(f)=J_f\,\bm{I}$, where $J_f=\int [f'(\xi)]^2/f(\xi)\,d\xi$ is the scalar Fisher information per observation. In this case, we then have $\tilde x_{it}=x_{it}-\bar x_i$ and
\begin{align}
 \mathcal{S}(\bm{x},f)
 = J_f\sum_{i,t}(x_{it}-\bar x_i)^2
 = J_f\,V(\bm{x}).
\end{align}
For Gaussian noise we have that $J_f=\sigma_\varepsilon^{-2}$, which is the smallest Fisher information for any density of variance $\sigma_\varepsilon^2$. To see this, consider estimating the location $\mu$ from a single draw $y=\mu+\varepsilon$ with $\varepsilon\sim f$. The draw itself, $\hat\mu=y$, is an unbiased estimator with variance $\sigma_\varepsilon^2$, so the Cram\'er--Rao inequality $\mathrm{Var}(\hat\mu)\geq 1/J_f$ requires that $\sigma_\varepsilon^2\geq 1/J_f$, that is, $J_f\geq\sigma_\varepsilon^{-2}$ for every $f$ of variance $\sigma_\varepsilon^2$. For Gaussian noise the score is $-\varepsilon/\sigma_\varepsilon^2$, so $J_f=\langle\varepsilon^2\rangle/\sigma_\varepsilon^4=\sigma_\varepsilon^{-2}$. Thus, since the sample mean of Gaussian draws is an unbiased location estimator, it saturates the Cram\'er--Rao bound and $\sigma_\varepsilon^{-2}$ is the smallest Fisher information among all densities of variance $\sigma_\varepsilon^2$~\cite{Cover2006Elements}. Evaluating $\delta\beta_{\min}=z/\sqrt{J_fV}$ at the Gaussian value is therefore conservative, as it is the largest limit compatible with the observed noise variance.

\clearpage
\section{Dynamical regimes of the resolution limit}\label{app:Hurst}

\subsection{Stationary deviations}

When the deviations are stationary, their covariance depends only on the lag, $\Sigma_{tt'}=\gamma(|t-t'|)$ with $\gamma(h)=\langle\xi_{it}\xi_{i,t+h}\rangle$ the autocovariance function, so that $\sigma_\xi^2=\gamma(0)$ is the marginal variance and $\tau=\sum_{h=1}^{\infty}\gamma(h)/\gamma(0)$ the correlation time. From the Szeg\H{o} limit theorem~\cite{Gray2006Toeplitz}, for $T\gg\tau$ the Toeplitz matrix $\bm{\Sigma}$ is asymptotically equivalent to a circulant matrix, and is therefore approximately diagonalized by the unitary discrete Fourier matrix $\bm{F}$ such that $F_{\omega t}=T^{-1/2}e^{-i\omega t}$ with $\omega=2\pi k/T$. The eigenvalues of $\bm{\Sigma}$ in this basis are given by the spectral density $P(\omega)=\sum_{h=-\infty}^{\infty}\gamma(h)e^{-i\omega h}$, so that Fourier inversion gives $\bm{\Sigma}\simeq\bm{F}^\dagger\,\mathrm{diag}\{P(\omega)\}\,\bm{F}$. Since $\bm{F}^\dagger\bm{F}=\bm{I}$, any quadratic form in $\bm{\Sigma}^{-1}$ can be evaluated in the Fourier basis,
\begin{align}
 \bm{u}^\top\bm{\Sigma}^{-1}\bm{u}
 \;\simeq\;\frac{1}{T}\sum_{\omega}\frac{|U(\omega)|^2}{P(\omega)},
 \qquad U(\omega)=\sqrt{T}\,(\bm{F}\bm{u})_\omega .
\end{align}

Consider the demeaned trend of a steadily growing system, $u_t=g(t-\bar t)$, with $t=0,\dots,T-1$ and $\bar t=(T-1)/2$. Its discrete Fourier coefficients at $\omega_k=2\pi k/T$ are, for $k\neq 0$ (the $k=0$ coefficient vanishes by demeaning),
\begin{align}
 U(\omega_k)=g\sum_{t=0}^{T-1}t\,e^{-i\omega_k t}=\frac{gT}{e^{-i\omega_k}-1},
 \qquad
 |U(\omega_k)|^2=\frac{g^2T^2}{4\sin^2(\omega_k/2)}\simeq\frac{g^2T^4}{4\pi^2k^2}.
\end{align}
As a check, the approximate spectrum satisfies Parseval's identity $T^{-1}\sum_{k=0}^{T-1}|U(\omega_k)|^2=\sum_{t=0}^{T-1}u_t^2$. On the right, we have $\sum_{t=0}^{T-1}u_t^2=g^2\sum_{t=0}^{T-1}(t-\bar t)^2=g^2\,T(T^2-1)/12$. Meanwhile, on the left we have that the $k=0$ term vanishes and the modes $k$ and $T-k$ are complex conjugates, so pairing them and using $|U(\omega_k)|^2\simeq g^2T^4/(4\pi^2k^2)$ for $1\leq k\ll T$ gives
\begin{align}
 \frac{1}{T}\sum_{k=1}^{T-1}|U(\omega_k)|^2
 \simeq\frac{1}{T}\cdot 2\sum_{k=1}^{\infty}\frac{g^2T^4}{4\pi^2k^2}
 =\frac{g^2T^3}{2\pi^2}\sum_{k=1}^{\infty}\frac{1}{k^2}
 =\frac{g^2T^3}{2\pi^2}\cdot\frac{\pi^2}{6}
 =\frac{g^2T^3}{12},
\end{align}
which agrees with the right side $g^2\,T(T^2-1)/12$ to leading order in $T$.

The fraction of the power at $|\omega_k|>\omega_c$, (equivalently, $|k|>K=\omega_cT/2\pi$), is then given by
\begin{align}
 \frac{\sum_{|k|>K}k^{-2}}{\sum_{k\neq0}k^{-2}}\simeq\frac{2/K}{\pi^2/3}=\frac{12}{\pi\,\omega_cT},
\end{align}
so with $\omega_c=1/\tau$ a fraction $12\tau/\pi T$ of the power lies at frequencies above $1/\tau$, which vanishes for $T\gg \tau$. For the remaining low frequencies, we can expand the spectral density to second order,
\begin{align}
 P(\omega)=\sum_h\gamma(h)\cos(\omega h)=P(0)\left[1-\tfrac{1}{2}\omega^2\tau_2^2+O(\omega^4)\right],
 \qquad
 \tau_2^2=\frac{\sum_h h^2\gamma(h)}{P(0)},
\end{align}
with $P(0)=\sum_h\gamma(h)=\sigma_\xi^2(1+2\tau)$ the long-run variance and $\tau_2$ finite when $\gamma(h)$ decays faster than $h^{-3}$. Splitting the Fourier sum at $K=T/2\pi\tau$ and inserting $|U(\omega_k)|^2\simeq g^2T^4/(4\pi^2k^2)$, $\omega_k=2\pi k/T$, and the expansion of $P(\omega)$, we have
\begin{align}
 \bm{u}^\top\bm{\Sigma}^{-1}\bm{u}
 &\simeq\frac{1}{T}\sum_{|k|\leq K}\frac{|U(\omega_k)|^2}{P(0)}
 -\frac{\tau_2^2}{2T}\sum_{|k|\leq K}\frac{\omega_k^2\,|U(\omega_k)|^2}{P(0)}
 +\frac{1}{T}\sum_{|k|>K}\frac{|U(\omega_k)|^2}{P(\omega_k)}
 \nonumber\\
 &=\frac{1}{P(0)}\left[\frac{g^2T^3}{12}-\frac{g^2T^3}{2\pi^2}\sum_{k>K}\frac{1}{k^2}\right]
 -\frac{\tau_2^2}{2T\,P(0)}\cdot g^2T^2\cdot 2K
 +\frac{1}{T}\sum_{|k|>K}\frac{g^2T^4}{4\pi^2k^2\,P(\omega_k)}
 \nonumber\\
 &=\frac{g^2T^3}{12\,P(0)}\left[1-\frac{6}{\pi^2K}-\frac{12\tau_2^2K}{T^2}+O\!\left(\frac{P(0)}{P_{\min}}\frac{1}{K}\right)\right]
 \nonumber\\
 &=\frac{g^2T^3}{12\,P(0)}\left[1+O\!\left(\frac{\tau}{T}\right)\right],
\end{align}
where the second term uses $\omega_k^2|U(\omega_k)|^2\simeq g^2T^2$ for every mode so that the sum over $2K$ modes gives $2Kg^2T^2$, and the third term is bounded by the high frequency contribution to the power, $\sum_{k>K}k^{-2}\simeq 1/K$, times $P(0)/P_{\min}$ with $P_{\min}=\min_\omega P(\omega)>0$. With $K=T/2\pi\tau$ every correction is $O(\tau/T)$, since the middle term is $12\tau_2^2/(2\pi\tau T)$ and $\tau_2\sim\tau$ for a short-memory process. Since $g^2T^3/12=\sum_t u_t^2$ and $P(0)=\sigma_\xi^2(1+2\tau)$, we finally have
\begin{align}
 \tilde{\bm{x}}_i^\top\bm{\Sigma}^{-1}\tilde{\bm{x}}_i
 =\frac{1}{\sigma_\xi^2(1+2\tau)}\sum_t(x_{it}-\bar x_i)^2\left[1+O(\tau/T)\right].
\end{align}
Summing over $m$ systems, we then have
\begin{align}
 \mathcal{S}\simeq\frac{V(\bm{x})}{\sigma_\xi^2(1+2\tau)},
 \qquad
 V(\bm{x})=\sum_{i,t}(x_{it}-\bar x_i)^2.
\end{align}

For sizes with Hurst exponent $H_x$, the within-system variation grows as $\sum_t(x_{it}-\bar x_i)^2\sim c_G\,T^{1+2H_x}$, where $c_G$ depends on the size dynamics. Specifically, we have the following common regimes for size dynamics:

\begin{itemize}

    \item \emph{Ballistic} ($H_x=1$, steady growth $x_{it}\approx x_{i0}+gt$): $\sum_t(x_{it}-\bar x_i)^2\approx g^2T^3/12$, such that $c_G=g^2/12$.

    \item \emph{Diffusive} ($H_x=1/2$, Gibrat growth with step variance $\sigma_g^2$): $\expec{\sum_t(x_{it}-\bar x_i)^2}=\sigma_g^2(T^2-1)/6$, such that $c_G=\sigma_g^2/6$.

    \item \emph{Confined} ($H_x=0$, fluctuations about a steady state with variance $\sigma_s^2$): $\sum_t(x_{it}-\bar x_i)^2\sim \sigma_s^2\,T$, such that $c_G=\sigma_s^2$.
    
\end{itemize}

In each case we have the general trend $\mathcal{S}\simeq m\,c_G\,T^{1+2H_x}/[\sigma_\xi^2(1+2\tau)]$, given in Eq.~\eqref{eq:Sstat}. With additive year effects $\theta_t$ in Eq.~\eqref{eq:model}, the two-way transformation $\tilde w_{it}=w_{it}-\bar w_{i\cdot}-\bar w_{\cdot t}+\bar w_{\cdot\cdot}$ eliminates both $c_i$ and $\theta_t$, such that the doubly demeaned ballistic regressor is $(g_i-\bar g)(t-\bar t)$ and every result above applies but with $g^2$ replaced by the across-system variance of growth rates $\tilde g^2=\frac{1}{m}\sum_i(g_i-\bar g)^2$.

\subsection{Nonstationary deviations}

The nonstationary regime is defined by a mean-square displacement that depends on the lag alone, such that $D(\ell)\equiv\langle(\xi_{i,t+\ell}-\xi_{it})^2\rangle=\sigma_\eta^2\,\ell^{2H_\xi}$ for every $t$. This is equivalent to the statement that the deviations $\bm{\xi}$ have stationary increments, with fractional Brownian motion being the Gaussian process with this property. Writing $X_a=\xi_{i,t+a}-\xi_{it}$, the identity $\langle X_aX_b\rangle=\tfrac{1}{2}[D(a)+D(b)-D(|a-b|)]$ follows from expanding $\langle(X_a-X_b)^2\rangle=D(|a-b|)$, so that the increments $\Delta\xi_{it}=\xi_{it}-\xi_{i,t-1}$ have covariance
\begin{align}
 \gamma_\eta(h)&=\langle\Delta\xi_{it}\,\Delta\xi_{i,t+h}\rangle \\
 &=\langle X_1X_{h+1}\rangle-\langle X_1X_h\rangle \\
 &=\tfrac{1}{2}\left[D(h+1)-2D(h)+D(h-1)\right] \\
 &=\tfrac{\sigma_\eta^2}{2}\left[(h+1)^{2H_\xi}-2h^{2H_\xi}+(h-1)^{2H_\xi}\right],
\end{align}
which depends only on the lag $h$ since the increments are a stationary process with marginal variance $\gamma_\eta(0)=\sigma_\eta^2$. For $H_\xi=1/2$ the bracket vanishes for all $h\geq1$ and the increments are independent. For $H_\xi>1/2$, $\gamma_\eta(h)$ is positive and decays as $h^{2H_\xi-2}$, such that the increments are positively correlated with long memory.

Letting $\Delta\bm{x}_i$ be the vector of $T-1$ size increments and $\bm{\Gamma}$ the $(T-1)\times(T-1)$ covariance matrix of the increments, $\Gamma_{tt'}=\gamma_\eta(|t-t'|)$, we have that $\mathcal{S}=\sum_i\Delta\bm{x}_i^\top\bm{\Gamma}^{-1}\Delta\bm{x}_i$ (differencing loses no information about $\beta$, since the level of each record is absorbed by its free prefactor $c_i$). Also let the size increments consist of drift plus Gibrat steps, $\Delta\bm{x}_i=g\bm{1}+\sigma_g\bm{\zeta}_i$, where the $\zeta_{it}$ are independent across $i$ and $t$ with $\langle\zeta_{it}\rangle=0$ and $\langle\zeta_{it}\zeta_{it'}\rangle=\delta_{tt'}$, so that $\langle\bm{\zeta}_i\rangle=\bm{0}$ and $\langle\bm{\zeta}_i\bm{\zeta}_i^\top\rangle=\bm{I}$. Expanding a single system's contribution to $\mathcal{S}$, we have
\begin{align}
 \Delta\bm{x}_i^\top\bm{\Gamma}^{-1}\Delta\bm{x}_i
 =g^2\,\bm{1}^\top\bm{\Gamma}^{-1}\bm{1}
 +2g\sigma_g\,\bm{1}^\top\bm{\Gamma}^{-1}\bm{\zeta}_i
 +\sigma_g^2\,\bm{\zeta}_i^\top\bm{\Gamma}^{-1}\bm{\zeta}_i .
\end{align}
Averaging over the steps, the cross term vanishes because $\langle\bm{\zeta}_i\rangle=\bm{0}$, and the last term gives $\langle\bm{\zeta}_i^\top\bm{\Gamma}^{-1}\bm{\zeta}_i\rangle=\sum_{tt'}(\bm{\Gamma}^{-1})_{tt'}\langle\zeta_{it}\zeta_{it'}\rangle=\sum_t(\bm{\Gamma}^{-1})_{tt}=\mathrm{tr}\,\bm{\Gamma}^{-1}$. Summing over the $m$ systems, we then have
\begin{align}
 \langle\mathcal{S}\rangle
 =m\left[g^2\,\bm{1}^\top\bm{\Gamma}^{-1}\bm{1}+\sigma_g^2\,\mathrm{tr}\,\bm{\Gamma}^{-1}\right].
\end{align}

For the first quadratic form in the brackets, we can apply the Cauchy--Schwarz inequality to $\bm{\Gamma}^{-1/2}\bm{1}$ and $\bm{\Gamma}^{1/2}\bm{1}$, with $\bm{\Gamma}^{1/2}$ the symmetric square root of $\bm{\Gamma}$, to get $(\bm{1}^\top\bm{1})^2\leq(\bm{1}^\top\bm{\Gamma}^{-1}\bm{1})(\bm{1}^\top\bm{\Gamma}\bm{1})$, with equality when $\bm{1}$ is an eigenvector of $\bm{\Gamma}$. Now, using $\bm{1}^\top\bm{\Gamma}\bm{1}=\mathrm{Var}\big(\sum_{t}\Delta\xi_{it}\big)=D(T-1)=\sigma_\eta^2(T-1)^{2H_\xi}$, we have that
\begin{align}
 \bm{1}^\top\bm{\Gamma}^{-1}\bm{1}\;\geq\;\frac{(T-1)^2}{\sigma_\eta^2(T-1)^{2H_\xi}}=\frac{(T-1)^{2(1-H_\xi)}}{\sigma_\eta^2},
\end{align}
with equality when $\bm{\Gamma}^{-1}\bm{1}\propto\bm{1}$ (i.e., for white noise increments) and equality up to a constant factor $c_{H}$ between $1$ and $1.02$ for all $H_\xi\in(0,1)$~\cite{Samarov1988On}, so that $\bm{1}^\top\bm{\Gamma}^{-1}\bm{1}=c_{H}(T-1)^{2(1-H_\xi)}/\sigma_\eta^2$.

For the second term, $\mathrm{tr}\,\bm{\Gamma}^{-1}$ is the sum of the eigenvalues of $\bm{\Gamma}^{-1}$, and the Szeg\H{o} theorem~\cite{Gray2006Toeplitz} states that for a Toeplitz matrix of size $n=T-1$ with symbol $P_\eta(\omega)$, the eigenvalues are asymptotically the values of the symbol on a grid of $n$ frequencies, so that for any continuous function $F$, $\frac{1}{n}\sum_{j}F(\lambda_j)\to\frac{1}{2\pi}\int_{-\pi}^{\pi}F(P_\eta(\omega))\,d\omega$. Using $F(\lambda)=1/\lambda$, we thus have
\begin{align}
 \mathrm{tr}\,\bm{\Gamma}^{-1}=\sum_j\frac{1}{\lambda_j}
 \simeq\frac{T-1}{2\pi}\int_{-\pi}^{\pi}\frac{d\omega}{P_\eta(\omega)}
 \equiv \kappa_H(T-1)/\sigma_\eta^2,
\end{align}
where $\kappa_H=\frac{\sigma_\eta^2}{2\pi}\int_{-\pi}^{\pi}\frac{d\omega}{P_\eta(\omega)}$ (equal to $1$ for white noise increments) with $P_\eta(\omega)=\sum_{h=-\infty}^{\infty}\gamma_\eta(h)e^{-i\omega h}$ the spectral density of the increments. Hence, we have
\begin{align}
 \langle\mathcal{S}\rangle
 \simeq\frac{m}{\sigma_\eta^2}\left[c_H\,g^2\,T^{2(1-H_\xi)}+\kappa_H\,\sigma_g^2\,T\right],
\end{align}
which for $H_\xi=1/2$ ($c_H=\kappa_H=1$) reduces to $m(g^2+\sigma_g^2)(T-1)/\sigma_\eta^2$. This result does not depend on $H_x$, since differencing reduces each time step to a single increment such that only the mean and variance of the increment enter the result.

\subsection{Within estimator in the nonstationary regime}

The within (fixed-effects) estimator of Eq.~\eqref{eq:FE} is given by $\hat\beta_{\FE}=\beta+\sum_{it}(x_{it}-\bar x_i)\,\xi_{it}/V(\bm{x})$, so its variance conditional on the sizes is
\begin{align}\label{eq:varFE}
 \mathrm{Var}(\hat\beta_{\FE}\mid\bm{x})
 = \frac{\sum_i\tilde{\bm{x}}_i^\top\bm{\Sigma}\,\tilde{\bm{x}}_i}{V(\bm{x})^2},
\end{align}
where $\tilde{\bm{x}}_i$ is the demeaned size vector as before. For a random walk, $\Sigma_{tt'}=\sigma_\eta^2\min(t,t')$, and with ballistic sizes this evaluates to $\mathrm{Var}(\hat\beta_{\FE})=\tfrac{6}{5}\,\sigma_\eta^2/(m g^2 T)$ (to leading order), which is only a constant factor $6/5$ above the limit $\sigma_\eta^2/[m g^2 (T-1)]$ of the first difference estimator for $H_x=1$. The mixed drift and Gibrat dynamics of the income panel ($H_x=0.88$) give a larger and slowly growing ratio, as seen in Table~\ref{tab:tracking}. Equation~\eqref{eq:varFE} evaluated on the actual size trajectories with the measured per-city step variances gives the random-walk curve of Fig.~\ref{fig:dynamics}(c). More generally, for deviations with Hurst exponent $H_\xi\geq 1/2$ the covariance scales as $\Sigma_{tt'}\sim\sigma_\eta^2\,T^{2H_\xi}$ for typical $t,t'$ in $[1,T]$, while each $\tilde x_{it}\sim T^{H_x}$. The numerator therefore scales as $T^2\cdot T^{2H_x}\cdot T^{2H_\xi}=T^{2+2H_x+2H_\xi}$ per system, and the denominator as $T^{2(1+2H_x)}$. Dividing these two factors gives
\begin{align}
 \mathrm{Var}(\hat\beta_{\FE})
 \;\sim\; T^{2+2H_x+2H_\xi-2(1+2H_x)}
 = T^{-2(H_x-H_\xi)}.
\end{align}
The resolution limit gives $\mathrm{Var}(\hat\beta_{\mathrm{GLS}})\sim T^{-2(1-H_\xi)}$ in this regime, so the within-estimator attains the asymptotic scaling of the resolution limit only when $H_x=1$ (ballistic sizes), in which case it sits a constant factor above it. For $H_x=H_\xi$ the variance is constant, and for $H_x<H_\xi$ it diverges, as in both cases the within fit regresses one nonstationary process on another.

\clearpage
\section{Empirical data and processing}
\label{app:data}

\subsection{Urban panels}
County-level personal income and population for 1969--2019 were obtained from the United States Bureau of Economic Analysis regional accounts (table CAINC1), with county-level GDP for 2001--2019 retrieved from table CAGDP1 and county-level employment by industry retrieved from table CAEMP25. Counties were mapped to metropolitan core-based statistical areas (CBSAs) under the 2019/2020 delineations, which were held fixed over the whole period of analysis, providing consistency to mitigate the sensitivities from boundary changes documented for existing scaling estimates~\cite{Arcaute2015Constructing,Cottineau2017Diverse}. We sum the member counties within each metro area to obtain balanced panels of $m=381$ metropolitan areas over $T=51$ years for population, income, and employment, and $T=19$ years for GDP. Income is given as nominal, and since outputs enter as logarithms the price level is a year effect absorbed by the two-way transformation described in Sec.~\ref{app:Hurst}. All panel estimates use two-way (city and year) demeaning with standard errors clustered by city, which allows arbitrary persistence of the deviations within each city's observation record. 

The exponents of Fig.~\ref{fig:empirical}(a) are:
\begin{itemize}
    \item income: $\hat\beta_{\CS}=1.071\pm 0.018$ vs $\hat\beta_{\FE}=1.013\pm 0.040$
    \item GDP: $\hat\beta_{\CS}=1.10 \pm 0.02$ vs $\hat\beta_{\FE}=0.94\pm 0.13$
    \item government employment: $\hat\beta_{\CS}=0.965\pm 0.030$ vs $\hat\beta_{\FE}=0.708\pm 0.054$
    \item farm employment: $\hat\beta_{\CS}=0.57\pm 0.03$ vs $\hat\beta_{\FE}=0.39\pm 0.12$
\end{itemize}
The errors for $\hat\beta_{\CS}$ are computed as two standard deviations over the per-period snapshots, and for $\hat\beta_{\FE}$ are computed using two standard errors. Including micropolitan areas (all $919$ CBSAs) gives income exponents of $\hat\beta_{\CS}=1.062 \pm 0.016$ and $\hat\beta_{\FE}=1.023\pm 0.035$, giving a similar gap. Notably, however, the sign of this gap is not universal across countries, as in Swedish wage data the within-city trajectories of the cities at the top of the urban hierarchy scale as steeply as the cross section~\cite{Keuschnigg2019Scaling}.

Total road length per CBSA and epoch (1900, 1910, \dots, 2010, 2015) was obtained from the historical road network reconstructions of~\cite{Burghardt2022Road,Burghardt2024Analyzing}, with matched census populations. Restricting to areas with positive population and road length in every epoch leaves a balanced panel of $m=743$ CBSAs over $13$ epochs. The cross-sectional exponent drifts from $0.93$ in 1900 to $0.73$ in 2015 (mean $0.84$, approximately the canonical infrastructure value~\cite{Bettencourt2007Growth}), so the snapshot bar in Fig.~\ref{fig:empirical}(a), $\pm 2$ standard deviations across the thirteen per-epoch fits, is $0.84\pm 0.15$. Within-city exponents with CBSA and epoch effects are $0.41\pm 0.06$ over the full century of data, and $0.06\pm 0.03$ over the final fifteen years, consistent with the durability of road networks. The dependence of the within-city exponent on window length for this indicator means the constant-$\beta$ model of Eq.~\eqref{eq:model} is an approximation here. Dropping the five-year 2015 epoch changes the century estimate by only $0.01$.

\begin{figure}[htb]
    \centering
    \includegraphics[width=0.6\textwidth]{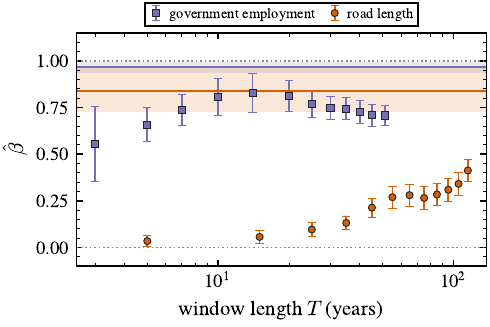}
    \caption{
    \textbf{Cross-sectional versus fixed-effects estimators for two additional urban indicators.}
    Within-city exponents versus window length ($2\sigma$ intervals, clustered by city) for government employment and total road length (markers). Dotted lines mark $\beta=0$ and $\beta=1$, while horizontal solid lines indicate the average cross-sectional exponents, with shading bands giving two standard deviations over all time periods.
    }
    \label{fig:other}
\end{figure}

Figure~\ref{fig:other} shows the within-city exponents of government employment and road length as a function of window length, to supplement the results of Fig.~\ref{fig:empirical}(a). The corresponding cross-sectional exponents are shown for comparison. For every window length the within-city exponent lies below the snapshot value, and for both indicators the gap persists as the window grows, implying that the bias is not an artifact of the window chosen.

\subsection{Metabolic scaling cross section}
From PanTHERIA~\cite{Jones2009PanTHERIA} we take all species with both basal metabolic rate (column 18-1) and the body mass measured alongside it (column 5-2), giving $n=573$ species spanning 5 orders of magnitude in mass. Orders with at least $20$ such species and at least five families are retained, leaving $442$ species in five orders (Rodentia, Chiroptera, Carnivora, Primates, Diprotodontia). The pooled fit, which demeans mass and metabolic rate within each order, gives slope $0.697\pm 0.014$, while per-order slopes are evaluated by least squares within each order, with variance $\hat\sigma^2/\mathrm{ss}_k$, $\mathrm{ss}_k$ being within-order variation of log mass. Each order plays the role of a system $i$ in Eq.~\eqref{eq:model} and its species the role of the observations $t$, so $\mathcal{S}_1=\mathrm{ss}_k/\hat\sigma^2$ and the limits of Sec.~\ref{app:groups} apply with $m=5$. All slopes are cross-sectional, implying (via Eq.~\eqref{eq:csbias}) that each estimates $\beta_k$ plus any prefactor--mass covariance.

With these variances the excess scatter statistic of Sec.~\ref{app:groups} gives $\hat\sigma_c=0.17$ versus $\sigma_{c,\min}=0.07$ and $\hat\sigma_\beta=0.10$ against $\sigma_{\beta,\min}=0.04$ (using the large-$m$ form $2\sqrt2/\sqrt m$, while the exact quantile $r_{\min}(5)$ of Sec.~\ref{app:groups} raises every limit by a factor $1.37$ without changing any conclusions). This implies that the metabolic scaling plausibly decomposes into separate prefactor and exponent subgroups defined by order. However, because related species have correlated residuals, we also recompute the order-level variances with cluster robust estimates (clustered by family). These cluster robust errors give $\hat\sigma_c=0.16$ versus $\sigma_{c,\min}=0.09$ (preserving the claim about prefactor subgroups), but result in $\hat\sigma_\beta\approx \sigma_{\beta,\min}=0.04$, suggesting a single universal exponent. (This change results largely from the large clustered exponent variance of Diprotodontia, whose $25$ species fall in ten families. Figure~\ref{fig:empirical}(b) plots the unclustered errors and limits.) The results are stable under different inclusion cutoffs (i.e., a species minimum count of $15$ or $30$, a family minimum count of $4$ or $6$).

\clearpage
\section{Deviation dynamics of the income panel}
\label{app:deviations}

\subsection{Hurst exponent fits}
For these calculations, the deviations $\bm{\xi}$ are computed as the two-way residuals $e_{it}$ of the full window within-city (fixed effects) fit, and the sizes are the demeaned log populations. For each we compute the mean-square displacement $\langle[e_{i,t+\ell}-e_{it}]^2\rangle$ over lags $\ell\in\{1,2,3,5,8,12,16,20\}$ years, averaged over cities and starting years, and fit $\propto\ell^{2H}$ by least squares [Fig.~\ref{fig:dynamics}(b)], giving $H_x=0.88$ and $H_\xi=0.47$. Fitting the residual displacement instead to $2\sigma_w^2+\sigma_\eta^2\ell$ (the sum of a stationary component of variance $\sigma_w^2$ and a random walk of step variance $\sigma_\eta^2$) gives $\sigma_w^2=2.0\times10^{-4}$ and $\sigma_\eta^2=5.1\times10^{-4}$ per year, meaning that the random walk dominates beyond $2\sigma_w^2/\sigma_\eta^2\approx0.8$~yr and thus every window analyzed is in the nonstationary regime. A pure power-law fit to this sum yields an apparent exponent slightly below $1/2$, which accounts for $H_\xi=0.47$. The size increments have across-city growth-rate dispersion $\tilde g=0.010$ and Gibrat step $\sigma_g=0.012$ per year. The yearly increments of the residuals are nearly uncorrelated (mean lag-one autocorrelation $0.11$ across cities), so the deviations are close to being a random walk with independent steps.

\subsection{Precision of the within fit}
Equation~\eqref{eq:varFE} gives the variance of the within fit for any covariance $\bm{\Sigma}$ of the deviations. For a random walk, $\Sigma_{tt'}=\sigma_{\eta,i}^2\min(t,t')$, and inserting the step variance $\hat\sigma_{\eta,i}^2$ measured from each city's increments gives the standard error of $\hat\beta_{\FE}$ over any window with no fitted parameter. Meanwhile, the resolution limit over the same window is given by $1/\sqrt{\mathcal{S}}$, with $\mathcal{S}=\sum_i\sum_t(\Delta x_{it})^2/\hat\sigma_{\eta,i}^2$ [Eq.~\eqref{eq:Snonstat}]. Table~\ref{tab:tracking} compares both with the observed standard error, computed with each city as a cluster, with the white noise prediction $\hat\sigma_\varepsilon/\sqrt{V}$ [Fig.~\ref{fig:dynamics}(c)]. The random-walk prediction matches the observed uncertainty to within $10\%$ for windows of $35$ years and longer and within $35\%$ at every window after $5$ years, while the white noise prediction lies below the observed value at $T=51$ by a substantial factor of $6.4$.

\begin{table}[H]
\caption{\label{tab:tracking}\textbf{Observed and predicted standard error of the within-city income exponent} ($\times10^{3}$) at each window length, for windows ending 2019. The random-walk rows use the measured step variances for individual cities and contain no fitted parameter, while the resolution limit row displays $1/\sqrt{\mathcal{S}}$ of Eq.~\eqref{eq:Snonstat}. The white noise row displays $\hat\sigma_\varepsilon/\sqrt{V}$, the standard error of the within fit for independent deviations, with $\hat\sigma_\varepsilon=0.066$ the residual standard deviation of the fit to the entire time window.}
\begin{ruledtabular}
\begin{tabular}{rcccccccccccc}
$T$ (yr) & 3 & 5 & 7 & 10 & 14 & 20 & 25 & 30 & 35 & 40 & 45 & 51 \\
\colrule
observed (cluster-robust) & 77.2 & 70.6 & 49.0 & 48.4 & 50.4 & 51.6 & 43.9 & 31.7 & 26.4 & 23.0 & 21.9 & 19.9 \\
random walk, within fit [Eq.~\eqref{eq:varFE}] & 128.3 & 75.5 & 63.4 & 62.2 & 56.5 & 40.6 & 32.7 & 27.6 & 24.5 & 22.1 & 20.2 & 18.3 \\
random walk, resolution limit & 75.1 & 51.2 & 42.7 & 36.7 & 30.1 & 22.5 & 19.4 & 16.9 & 14.6 & 13.1 & 11.5 & 9.8 \\
white noise & 265.8 & 119.4 & 72.6 & 45.7 & 29.7 & 16.3 & 11.3 & 8.2 & 6.2 & 4.9 & 3.9 & 3.1 \\
\end{tabular}
\end{ruledtabular}
\end{table}

\subsection{Resolution windows}
The nonstationary columns of Table~\ref{tab:ladder} are the windows at which the predicted and the observed standard errors of Table~\ref{tab:tracking} cross the level $\delta\beta/2$ (Sec.~\ref{app:KL}), which is the standard error at which a question with exponent separation $\delta\beta$ is resolvable. For competing theories with $\delta\beta=0.05$, these thresholds for the within fit are $34$ and $37$ years, whereas the resolution limit itself crosses at about $18$ years, since the within fit is not the GLS fit under a random walk (Sec.~\ref{app:Hurst}) so does not attain the resolution limit. (The closed form of Eq.~\eqref{eq:tstar}, which assumes every year carries the same information, gives a window of $10$ years using the panel's mean calibration. However, the actual windows ending in 2019 need longer because the size increments of the most recent decade are smaller and more uniform across cities, contributing much less information per year than earlier decades.) Extrapolating the random-walk law to $m=1$ (a single city) gives $762$, $1930$, and $12{,}970$ years for resolving the three questions. However, on these time scales the deviations cannot keep diffusing, and if information accumulates as in Eq.~\eqref{eq:Sstat} with $\sigma_\xi=0.15$ (the residual standard deviation about the annual cross sections) and $\tau=100$~yr (the relaxation time implied by the $0.61$ correlation between a city's 1969 and 2019 residuals), this shortens the single-city windows to about $130$, $460$, and $950$ years respectively (which is still beyond any measured panel record). Thus, resolving a scaling exponent (even just for determining nonlinearity $\beta\neq 1$) for this dataset is not possible for a single city without a much longer set of temporal observations.

\begin{table}[H]
\caption{\label{tab:ladder}\textbf{Resolution windows for the three exponent location questions.} Windows $T^{*}$ (in years) for a corpus ($m>1$) and for a single city ($m=1$) are presented for exponent separations $\delta\beta$ relevant to each question. For the stationary columns, we use Eq.~\eqref{eq:tstar} with $H_x=1$, $c_G=g^2/12$, and the parameters from the synthetic experiments ($g=0.015$, $\sigma_\xi=0.1$, $\tau=0$, Sec.~\ref{app:synthetic}). For the nonstationary columns, we show the random-walk prediction for the within fit on the income panel [Eq.~\eqref{eq:varFE}], next to the result for cluster robust standard errors ($3$~yr is the shortest window computed). The $m=1$ columns evaluate the same laws for a single city at the corpus calibration (an idealization, since a single record cannot remove year effects).}
\begin{ruledtabular}
\begin{tabular}{l@{\hspace{4pt}}c@{\hspace{6pt}}cc@{\hspace{6pt}}ccc}
 & & \multicolumn{2}{c}{synthetic, stationary} & \multicolumn{3}{c}{income, nonstationary} \\
Question & $\delta\beta$ & $m=300$ & $m=1$ & $m=381$ & observed & $m=1$ \\
\colrule
$\beta\neq 0$ & 1 & 1.9 & 13 & $\leq 2$ & $\leq 3$ & 762 \\
$\beta\neq 1$ & 0.15 & 6.8 & 46 & 5 & 4 & 1930 \\
$\beta_1$ vs $\beta_2$ & 0.05 & 14.2 & 95 & 34 & 37 & 12970 \\
\end{tabular}
\end{ruledtabular}
\end{table}

\clearpage
\section{Resolution limits for subgroups}
\label{app:groups}

\subsection{Reduction to per-system estimates}
Fitting Eq.~\eqref{eq:model} for a single system $i$ gives a least-squares slope of $\hat\beta_i=\sum_t\tilde x_{it}y_{it}/\sum_t\tilde x_{it}^2$, with $\tilde x_{it}=x_{it}-\bar x_i$. Substituting $y_{it}=c_i+\beta_ix_{it}+\xi_{it}$, the prefactor $c_i$ drops out because $\sum_t\tilde x_{it}=0$ and the term $\beta_ix_{it}$ contributes $\beta_i\sum_t\tilde x_{it}^2/\sum_t\tilde x_{it}^2=\beta_i$, so we can rewrite this as $\hat\beta_i=\beta_i+\nu_i$ with $\nu_i=\sum_t\tilde x_{it}\xi_{it}/\sum_t\tilde x_{it}^2$. The error $\nu_i$ is a linear combination of $i$'s deviations, and so has mean zero and is Gaussian when these deviations $\bm{\xi}_i$ are Gaussian, with variance
\begin{align}
 \mathrm{Var}(\nu_i)=\frac{\tilde{\bm{x}}_i^\top\Sigma\,\tilde{\bm{x}}_i}{(\tilde{\bm{x}}_i^\top\tilde{\bm{x}}_i)^2}.
\end{align}
This is just Eq.~\eqref{eq:varFE} for one system, and reduces to $\sigma_\xi^2/\sum_t\tilde x_{it}^2$ for white noise deviations. By the Cramér--Rao argument of Sec.~\ref{app:KL}, no estimator of $\beta_i$ from system $i$ has variance below $1/\mathcal{S}_1$, with $\mathcal{S}_1=\tilde{\bm{x}}_i^\top\Sigma^{-1}\tilde{\bm{x}}_i$. The GLS fit of Sec.~\ref{app:KL} attains the resolution limit, allowing us to take $v=1/\mathcal{S}_1$ as the variance of the per-system estimate. (For white noise deviations, the least-squares and GLS fits coincide and the two variances above are equal.) $\mathcal{S}_1$ is given by either Eq.~\eqref{eq:Sstat} or \eqref{eq:Snonstat} evaluated at $m=1$, which is $c_GT^{1+2H_x}/[\sigma_\xi^2(1+2\tau)]$ for stationary deviations and $\sum_t(\Delta x_{it})^2/\sigma_\eta^2$ for nonstationary deviations. (In the nonstationary regime the fit should be applied to the first differences, since this increment estimator is the GLS fit that attains $\mathcal{S}_1$ while the standard fit does not.)

Now let the exponents be $\beta_i=\beta+\delta_i$, with $\delta_i$ drawn independently from some distribution $G$ with mean zero and variance $\sigma_\beta^2$, of otherwise arbitrary shape. Then $\hat\beta_i=\beta+\delta_i+\nu_i$ is the sum of two independent terms, so its variance is $v+\sigma_\beta^2$ when the exponents vary and $v$ when they do not. All information about $\sigma_\beta$ can be found from the $m$ independent values $\hat\beta_i$, since the exponent of system $i$ enters the likelihood only through system $i$'s measurements, and once the estimate $\hat\beta_i$ has been extracted the remaining statistics do not depend on $\delta_i$. The common mean $\beta$ is the same whether or not the exponents vary, so we work with the centered values $u_i=\hat\beta_i-\beta$. Let $\phi$ be the density of the estimation error $\nu_i$. For the efficient single-system estimator this is $\mathcal{N}(0,v)$ exactly under Gaussian deviations and asymptotically follows this distribution under a general deviation density $f$, with $v=1/\mathcal{S}_1$ and $\mathcal{S}_1=\tilde{\bm{x}}_i^\top\mathcal{I}(f)\tilde{\bm{x}}_i$ as in Sec.~\ref{app:KL}. Thus the result depends on the noise density $f$ only through the value $v$. We will keep $\phi$ as a general density below, and specialize to the Gaussian case only where it is needed. The density of $u_i$ is then equal to $\phi$ when the exponents are equal and equal to the convolution $p(u)=\int\phi(u-\delta)\,dG(\delta)$ otherwise. Resolving $\sigma_\beta>0$ is then reduced to the problem of distinguishing $p$ from $\phi$ (in KL divergence) using $m$ samples.

\subsection{Resolution limit for exponent subgroups}
Expanding $\phi(u-\delta)$ in powers of $\delta$ and integrating over $G$ (whose first moment vanishes by definition) gives
\begin{align}
 p(u)=\phi(u)+\frac{\sigma_\beta^2}{2}\,\phi''(u)-\frac{\mu_3}{6}\,\phi'''(u)+O(\sigma_\beta^4),
\end{align}
with $\mu_3$ the third moment of $G$. Writing $p=\phi(1+h)$, the KL divergence analogous to Eq.~\ref{eq:klmain} is given by
\begin{align}
 \mathrm{KL}(p\,\|\,\phi)=\int p\ln\frac{p}{\phi}=\int\phi\,(1+h)\ln(1+h)=\int\phi\,h+\frac{1}{2}\int\phi\,h^2+O(h^3).
\end{align}
The first term vanishes, because $\int\phi\,h=\int(p-\phi)=0$. The leading part of $h$ is thus $\tfrac{\sigma_\beta^2}{2}\phi''/\phi$, so for any noise density we have
\begin{align}
 \mathrm{KL}(p\,\|\,\phi)=\frac{\sigma_\beta^4}{8}\,J_2(\phi)+\dots,
\end{align}
with
\begin{align}
J_2(\phi)=\int\frac{[\phi''(u)]^2}{\phi(u)}\,du.    
\end{align}
(The $\mu_3$ term does not contribute at this order because its cross term with $\phi''$ integrates to zero for any symmetric $\phi$, and is of order $O(\sigma_\beta^5)$ otherwise.) 

For the noise density $\phi=\mathcal{N}(0,v)$, we have that $\phi''/\phi=u^2/v^2-1/v$ so that $J_2=2/v^2$, giving
\begin{align}
 \mathrm{KL}(p\,\|\,\phi)=\frac{\sigma_\beta^4}{4v^2}+\dots=\frac{\kappa^2}{4}+O(\kappa^3),
\end{align}
with $\kappa=\sigma_\beta^2/v$. (For Gaussian $G$, this result can be checked against the closed form of $\mathrm{KL}(\mathcal{N}(0,v+\sigma_\beta^2)\,\|\,\mathcal{N}(0,v))=\tfrac12[\kappa-\ln(1+\kappa)]$.) Over $m$ independent systems the KL divergence adds to give $m\kappa^2/4$, so that the criterion of Sec.~\ref{app:KL} then gives
\begin{align}\label{eq:kappamin}
 \frac{m\kappa^2}{4}\geq 2
 \quad\Longleftrightarrow\quad
 \sigma_\beta^2\geq\frac{2\sqrt{2}}{\sqrt{m}}\,v=\frac{2\sqrt{2}}{\sqrt{m}\,\mathcal{S}_1}.
\end{align}
Substituting $\mathcal{S}_1$ for the stationary and nonstationary deviation regimes gives the exponent limits of Eqs.~\eqref{eq:sigbetamin} and \eqref{eq:sigbetamin2}. For non-Gaussian $\phi$, the same derivation gives $\sigma_\beta^2\geq4/\sqrt{m\,J_2(\phi)}$.

\subsection{Resolution limit for prefactor subgroups}
Averaging Eq.~\eqref{eq:model} over the record of system $i$ and inserting the pooled exponent, the fitted prefactor of system $i$ is given by $\hat c_i=\bar y_i-\hat\beta_{\FE}\bar x_i=c_i+\bar\xi_i-(\hat\beta_{\FE}-\beta)\bar x_i$. For stationary deviations, the second term is the temporal mean of $T$ correlated draws, with variance $\sigma_\xi^2(1+2\tau)/T$ and $1+2\tau=\sum_{\ell}\rho(\ell)$ the sum of the deviation autocorrelations. Meanwhile the third term is a common error proportional to $\bar x_i$, which can be removed by regressing the prefactors $c_i$ on $\bar x_i$ and testing the residuals. The individual system information for the prefactor $c_i$ is therefore $\mathcal{S}_1=T/[\sigma_\xi^2(1+2\tau)]$ regardless of the size dynamics, since every observation is equally informative about a given system, and Eq.~\eqref{eq:kappamin} gives
\begin{align}
\sigma_{c,\min}^2=2\sqrt{2}\,\sigma_\xi^2(1+2\tau)/(\sqrt{m}\,T),    
\end{align}
which is the prefactor subgroup resolution limit of Eq.~\eqref{eq:sigbetamin}. For nonstationary deviations, the temporal mean $\bar\xi_i$ of a random walk has variance $\sigma_\eta^2\sum_{t,t'}\min(t,t')/T^2\simeq\sigma_\eta^2T/3$, which grows with the time window $T$, implying that a system's time-averaged prefactor is not a fixed property of the system that becomes more precise with additional observation. In this regime it is thus more relevant to ask whether $c_i+\xi_{it}$ at a given time step are more dispersed than permitted by a single prefactor. (This is why Fig.~\ref{fig:phase} draws only the exponent boundaries.)

\subsection{Attainment}
Given per-system estimates $\hat\beta_i$ with known noise variances $v_i$, the standard heterogeneity statistic of meta-analysis~\cite{Hedges2001The}, which here we call $Q$, weights each squared deviation from the precision-weighted mean by its own precision $1/v_i$, thus
\begin{align}
 Q=\sum_{i=1}^m\frac{(\hat\beta_i-\bar\beta_w)^2}{v_i},
 \qquad
 \bar\beta_w=\frac{\sum_i\hat\beta_i/v_i}{\sum_i 1/v_i}.
\end{align}
When all exponents $\beta_i$ are equal, the standardized values $(\hat\beta_i-\beta)/\sqrt{v_i}$ are $m$ independent standard normals such that $Q\sim\chi^2_{m-1}$, with mean $m-1$ and variance $2(m-1)$~\cite{Hedges2001The,Swamy1970Efficient,Pesaran2008Testing}. We then have that $[Q-(m-1)]/\sqrt{2(m-1)}$, which we call the excess scatter score, is therefore a standard normal under the single exponent case. When the exponents are instead spread with variance $\sigma_\beta^2$, the variance of $\hat\beta_i$ about the common mean becomes $v_i+\sigma_\beta^2$, and taking the expectation of $Q$ term by term with $w_i=1/v_i$ gives
\begin{align}
\expec{Q}=\sum_iw_i(v_i+\sigma_\beta^2)-\sum_iw_i^2(v_i+\sigma_\beta^2)/\sum_iw_i=(m-1)+\sigma_\beta^2\left[\sum_iw_i-\sum_iw_i^2/\sum_iw_i\right].    
\end{align}
Solving this for the spread that reproduces the observed $Q$, we have
\begin{align}
 \hat\sigma_\beta^2=\frac{Q-(m-1)}{\sum_iw_i-\sum_iw_i^2/\sum_iw_i},
\end{align}
which is the estimate $\hat\sigma_\beta$ ($\hat\sigma_c$, for the prefactors) in Sec.~\ref{app:data}. 

Under a single exponent $\beta_i=\beta$, $Q\sim \chi^2_{m-1}$, so $Z=[Q-(m-1)]/\sqrt{2(m-1)}$ has mean zero and unit variance. Thus, following the two standard error criterion of Sec.~\ref{app:KL}, we can determine that a spread $\sigma_\beta$ exists for $Z>2$. To see that this happens exactly at the limit of Eq.~\eqref{eq:kappamin}, we can take equal noise variances $v_i=v$, for which the expectation derived above reduces to $\expec{Q}=(m-1)+\sigma_\beta^2(m-1)/v=(m-1)(1+\kappa)$ with $\kappa=\sigma_\beta^2/v$. A spread $\kappa$ therefore shifts the mean score to
\begin{align}
 \expec{Z}=\frac{(m-1)\kappa}{\sqrt{2(m-1)}}=\kappa\sqrt{\frac{m-1}{2}},
\end{align}
which at $\kappa=2\sqrt{2}/\sqrt{m}$ equals $2\sqrt{(m-1)/m}\simeq2$. At the resolution limit the value of $Z$ is therefore distributed with unit variance and mean $2$, so it exceeds the threshold $Z^\ast=2$ in half of simulations. This implies that this test can resolve an exponent spread (e.g. exponent subgroups) with probability one half exactly at the resolution limit, which is the same criterion as in Sec.~\ref{app:KL}, meaning that this excess scatter test achieves the resolution limit. 

At small $m$, the $\chi^2_{m-1}$ null is skewed and $Z$ is not well approximated by a normal distribution, so the simulations and the boundaries of Fig.~\ref{fig:phase} use the exact null model quantiles in place of $2\sqrt{2}/\sqrt{m}$, replacing it by
\begin{align}
 r_{\min}(m)=\frac{F^{-1}_{\chi^2_{m-1}}(\Phi(2))}{\mathrm{med}(\chi^2_{m-1})}-1,
\end{align}
which converges to $r_{\min}(m)\to2\sqrt{2}/\sqrt{m}$ for large $m$.

\clearpage
\section{Numerical verification of analytical results}
\label{app:synthetic}

\begin{figure}[b]
    \centering
    \vspace{20pt}
    \includegraphics[width=0.5\textwidth]{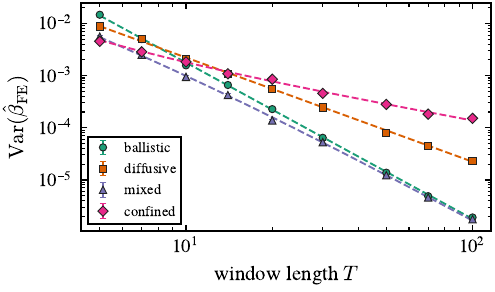}
    \caption{
    \textbf{The $T^{-(1+2H_x)}$ family of variance laws.}
    Variance of $\hat\beta_{\FE}$ versus window length $T$ for ballistic ($H_x=1$, $g=0.015$), diffusive ($H_x=1/2$, $\sigma_g=0.03$), mixed (both Gibrat and drift terms), and confined ($H_x=0$, $\sigma_s=0.05$) size dynamics, with $\sigma_\xi=0.1$. Dashed lines are the theoretical variance $\sigma_\xi^2/V$ of Sec.~\ref{app:Hurst}, and error bars are one standard error.
    }
    \label{fig:synthetic}
\end{figure}

\begin{figure}
    \centering
    \includegraphics[width=0.5\textwidth]{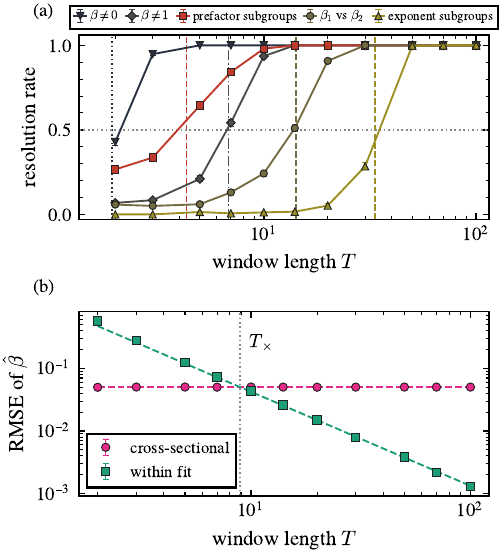}
    \caption{
    \textbf{Numerical verification of resolution rates and estimator errors.}
    (a)~Fraction of simulations in which each question is resolved versus $T$, for $\beta\neq 0$ ($\delta\beta=1$), $\beta\neq 1$ ($\delta\beta=0.15$), $\beta_1$ vs $\beta_2$ ($\delta\beta=0.05$), prefactor subgroups ($\sigma_c=0.02$), and exponent subgroups ($\sigma_\beta=0.05$). Vertical lines mark the predicted boundaries $T^{*}$ computed using Eq.~\eqref{eq:tstar} and Eq.~\eqref{eq:sigbetamin}, and the horizontal dotted line marks $50\%$ of simulations in which the question is resolved.
    (b)~RMSE of $\hat\beta_{\CS}$ and $\hat\beta_{\FE}$ with a prefactor--size slope of $b=0.05$. Dashed lines indicate the theoretical errors, which cross at $T_\times\approx 8.9$ (dotted vertical line). We set $m=300$, $g=0.015$, $\sigma_\xi=0.1$, and run $500$ simulations per point in both panels.
    }
    \label{fig:detection}
\end{figure}

In all simulations we sample from Eq.~\eqref{eq:model} with initial sizes log-uniform over $[10^{4.5},10^{7.5}]$. Figures~\ref{fig:synthetic} and \ref{fig:detection} use a drift $g=0.015$ per period, Gaussian white noise deviations $\sigma_\xi=0.1$ (i.e., $\tau=0$), Gibrat steps $\sigma_g=0.03$ where sizes are nonstationary, $m=300$, and $500$ repetitions per point. Meanwhile, Fig.~\ref{fig:phase} uses the measured income calibration described in the corresponding caption. Under the two standard error criterion of Sec.~\ref{app:KL}, an exponent location question counts as resolved when $|\hat\beta_{\FE}-\beta_0|>2\,\hat\sigma_\xi/\sqrt{V}$. We set $(\beta,\beta_0)=(1,0)$, $(1.15,1)$, and $(1.05,1)$ for the three tasks. Meanwhile, a subgroup question (for exponents or prefactors) is resolved when the statistic $Q$ of Sec.~\ref{app:groups} exceeds the $\chi^2_{m-1}$ quantile at level $\Phi(2)$.

Figure~\ref{fig:synthetic} scans over a range $T\in\{5,\dots,100\}$ for ballistic ($H_x=1$), diffusive ($H_x=1/2$), mixed (drift and Gibrat steps together), and confined ($H_x=0$) sizes. The simulated variance of $\hat\beta_{\FE}$ follows $\sigma_\xi^2/V$ with the $c_G$ factors found in Sec.~\ref{app:Hurst} and slopes $T^{-3},T^{-2},T^{-1}$ for $H_x=1,1/2,0$ respectively, with no additional fitted parameters.

Figure~\ref{fig:detection}(a) scans over a range $T\in\{2,\dots,100\}$ for the three exponent location estimation tasks as well as for the prefactor ($\sigma_c=0.02$) and exponent ($\sigma_\beta=0.05$) subgroup resolution tasks. We can see that the fraction of simulations in which each question is resolved passes through one half at the predicted $T$ values. Figure~\ref{fig:detection}(b) compares the root-mean-square errors of $\hat\beta_{\CS}$ and $\hat\beta_{\FE}$ under a prefactor--size slope $b=0.05$, where we can see that the snapshot error tends to the bias $|b|$ [Eq.~\eqref{eq:csbias}] while the panel error falls as $T^{-3/2}$. The two errors cross at roughly $T_\times=(12\sigma_\xi^2/mg^2b^2)^{1/3}\approx 8.9$, which is $T^{*}$ of Eq.~\eqref{eq:tstar} evaluated at $\delta\beta=2b$.

\subsection{Resolution diagram}
Figure~\ref{fig:phase} scans over nine corpus sizes $m\in\{10,\dots,1000\}$ and fifteen time windows $T\in\{2,\dots,200\}$ with $200$ simulations per cell, and the plotted points are the windows at which the fraction of times the corresponding exponent location question is resolved crosses one half. In the stationary panel (top) they agree with the predicted boundaries to within $6\%$ for the $\beta_1\neq \beta_2$ and $\beta\neq 1$ tasks, $4\%$ for exponent subgroups at every $m$, and within $12\%$ for $\beta\neq 0$. Meanwhile, the nonstationary panel samples the deviations as a random walk of step $\sigma_\eta=0.0226$ on sizes with drift $\tilde g=0.010$ and Gibrat step size $\sigma_g=0.012$, which correspond to the income calibration of Sec.~\ref{app:deviations}. Each exponent question is then resolved using the increment estimators of Secs.~\ref{app:Hurst} and \ref{app:groups}. Since an increment fit over a window of $T$ years uses $T-1$ independent steps, each of expected squared size $(\tilde g^2+\sigma_g^2)$ and noise variance $\sigma_\eta^2$, the boundaries are given by $T^{*}=1+4\sigma_\eta^2/[(\tilde g^2+\sigma_g^2)m\,\delta\beta^2]$ for an exponent location question [Eq.~\eqref{eq:tstar}] and $T=1+r_{\min}(m)\,\sigma_\eta^2/[(\tilde g^2+\sigma_g^2)\sigma_\beta^2]$ for the exponent subgroup question. These are evaluated at $\sigma_\beta=0.1$, the scale of the measured income resolution limit. The measured crossings agree with the predictions to within $12\%$ for $\beta_1\neq\beta_2$, $13\%$ for $\beta\neq 1$, and $7\%$ for exponent subgroups. Prefactor boundaries are not drawn in either panel (see Sec.~\ref{app:groups}).

\subsection{Scaling discrepancies and the choice of $\delta\beta$}
The separation $\delta\beta=1$ is chosen to match the distance $\delta\beta=1-0$ between no scaling ($\beta=0$) and linear scaling ($\beta=1$). Meanwhile, the separation $\delta\beta=0.15$ is chosen to match the distance $\delta\beta=1.15-1$ between linear scaling ($\beta=1$) and the standard superlinear scaling ($\beta=1.15$) of urban scaling theories. Finally, the separation $\delta\beta=0.05$ is chosen to capture the typical discrepancies between mechanistic predictions as well as between these predictions and empirical fits. The urban scaling theory of Ref.~\cite{Bettencourt2013The} fixes the socioeconomic exponent at $\beta=1+\delta$ with $\delta=H_n/[D(D+H_n)]=1/6$ for a two-dimensional city (with $H_n$ the network dimension of Ref.~\cite{Bettencourt2013The}, unrelated to the Hurst exponents above), i.e.\ $\beta\simeq1.17$, while finding an observed $\delta\simeq0.15$ for wages and inventions. A wage exponent of $1.12$ is also found in Ref.~\cite{Bettencourt2007Growth}. The models of Refs.~\cite{Ribeiro2017A,GomezLievano2017Explaining} instead make $\beta$ a continuous function of the range of social interactions relative to the city's fractal dimension, or of the number of complementary factors required by some phenomenon, such that exponents differing by a few hundredths correspond to distinct parameter values. The same scale separates existing mechanistic predictions for metabolic scaling. Geometric surface area arguments give $\beta=2/3$~\cite{White2003Mammalian} for the scaling of metabolic rate with body mass, while the network theory of Ref.~\cite{West1997A} gives $3/4$, and existing analyses of mammalian data are found to support one or the other by margins of a few hundredths. For example, a range of $0.65$--$0.68$ is identified in Ref.~\cite{White2003Mammalian}, while intervals including $3/4$ but excluding $2/3$ are found in Ref.~\cite{Savage2004The}. The common slope $0.697\pm0.014$ of Sec.~\ref{app:data} lies between the two values, $0.03$ from one and $0.05$ from the other. Distinguishing among these different theories and empirical results therefore requires resolving differences of order $\delta\beta \approx 0.05$.

\end{document}